\pdfoutput=1
\documentclass[journal,a4paper,twocolumn,10pt]{IEEEtran}

\usepackage{amsmath}
\usepackage[british,UKenglish,USenglish,american]{babel}
\usepackage{xfrac} % for \sfrac{}{}
\usepackage{bm} % for bold greek letters
\usepackage{amsfonts}
\usepackage{amssymb}
\usepackage{color}
\usepackage{graphicx}
\usepackage{graphics}
\usepackage{algorithm}% http://ctan.org/pkg/algorithms
\usepackage{algpseudocode}% http://ctan.org/pkg/algorithmicx
\usepackage{acro} 
\usepackage{array} % for vertical centering in a table
\usepackage{pbox}
\usepackage{subcaption}
\usepackage{hyperref}
\usepackage{cite}
\usepackage{dblfloatfix}% To enable figures at the bottom of page
\usepackage{placeins}

\usepackage{xcolor}
\usepackage[switch]{lineno}

\usepackage{bm}

\usepackage{pgfplots}
\usepackage{tikz}
\usetikzlibrary{shapes.geometric}
\usetikzlibrary{arrows.meta}
\usetikzlibrary{external}
\usetikzlibrary{plotmarks}
\usepackage{algorithm}
\usepackage{algorithmicx}
\algdef{SE}[DOWHILE]{Do}{doWhile}{\algorithmicdo}[1]{\algorithmicwhile\ #1}%

\newcommand{\equref}[1]{(\ref{#1})}

\DeclareMathOperator{\Exp}{E}

\DeclareMathOperator{\cov}{Cov}
\DeclareMathOperator{\diag}{diag}

\DeclareMathOperator{\mse}{MSE}

\DeclareMathOperator{\PT}{\mathrm{PT}}
\DeclareMathOperator{\snr}{\mathrm{SNR}}
\DeclareMathOperator{\dB}{\mathrm{dB}}

\newcommand{\mbf}[1]{\mathbf{#1}}
\newcommand{\mcal}[1]{\mathcal{#1}}

\newcommand{\mbsf}[1]{\boldsymbol{\mathsf{#1}}}
\newcommand{\mbb}[1]{\mathbb{#1}}

\newcommand{\av}{\mbf{a}}

\newcommand{\ev}{\mathbf{e}}

\newcommand{\hv}{\mathbf{h}}

\newcommand{\rv}{\mathbf{r}}

\newcommand{\uv}{\mathbf{u}}
\newcommand{\vv}{\mathbf{v}}
\newcommand{\wv}{\mathbf{w}}
\newcommand{\xv}{\mathbf{x}}
\newcommand{\yv}{\mathbf{y}}
\newcommand{\zv}{\mathbf{z}}

\newcommand{\hvec}{\vec{\hv}}

\newcommand{\rvec}{\vec{\rv}}

\newcommand{\uvec}{\vec{\uv}}
\newcommand{\vvec}{\vec{\vv}}
\newcommand{\wvec}{\vec{\wv}}
\newcommand{\xvec}{\vec{\xv}}
\newcommand{\yvec}{\vec{\yv}}
\newcommand{\zvec}{\vec{\zv}}

\newcommand{\Av}{\mbf{A}}

\newcommand{\Cv}{\mathbf{C}}

\newcommand{\Ev}{\mathbf{E}}

\newcommand{\Gv}{\mathbf{G}}

\newcommand{\Iv}{\mathbf{I}}

\newcommand{\Pv}{\mathbf{P}}
\newcommand{\Qv}{\mathbf{Q}}

\newcommand{\Tv}{\mathbf{T}}

\newcommand{\Wv}{\mathbf{W}}
\newcommand{\Xv}{\mathbf{X}}
\newcommand{\Yv}{\mathbf{Y}}

\newcommand{\zerov}{\mathbf{0}}

\newcommand{\vs}{\mathsf{v}}
\newcommand{\ws}{\mathsf{w}}
\newcommand{\xs}{\mathsf{x}}

\newcommand{\uvs}{\mbsf{u}}
\newcommand{\vvs}{\mbsf{v}}
\newcommand{\wvs}{\mbsf{w}}
\newcommand{\xvs}{\mbsf{x}}
\newcommand{\yvs}{\mbsf{y}}

\newcommand{\uvecs}{\vec{\uvs}}
\newcommand{\vvecs}{\vec{\vvs}}
\newcommand{\wvecs}{\vec{\wvs}}
\newcommand{\xvecs}{\vec{\xvs}}
\newcommand{\yvecs}{\vec{\yvs}}

\newcommand{\Avs}{\mbsf{A}}

\newcommand{\Wvs}{\mbsf{W}}
\newcommand{\Xvs}{\mbsf{X}}
\newcommand{\Yvs}{\mbsf{Y}}

\newcommand{\eps}{\epsilon}
\newcommand{\vareps}{\varepsilon}

\newcommand{\sw}{\sigma_{\ws}}

\newcommand{\Sx}{\boldsymbol{\Sigma}_{\xvecs}}
\newcommand{\Sw}{\boldsymbol{\Sigma}_{\wvecs}}
\newcommand{\Svv}{\boldsymbol{\Sigma}_{\vvecs}}
\newcommand{\Su}{\boldsymbol{\Sigma}_{\uvecs}}
\newcommand{\Sxt}{\boldsymbol{\Sigma}_{\tilde{\xvecs}}}
\newcommand{\Swt}{\boldsymbol{\Sigma}_{\tilde{\wvecs}}}
\newcommand{\Svvt}{\boldsymbol{\Sigma}_{\tilde{\vvecs}}}
\newcommand{\Srr}{\boldsymbol{\Sigma}_{\vec{r}}}
\newcommand{\Evec}{\vec{\Ev}}

\newcommand{\xhat}{\hat{x}}
\newcommand{\xvhat}{\hat{\xv}}
\newcommand{\xvechat}{\hat{\vec{\xv}}}

\newcommand{\Xvhat}{\hat{\Xv}}

\newcommand{\be}{\begin{equation}}
\newcommand{\ee}{\end{equation}}
\newcommand{\lef}{\left}
\newcommand{\rig}{\right}
\newcommand{\tol}{\mathrm{tol}}

\def\*#1{\mathbf{#1}}

\DeclareAcronym{cs}{
	short = CS,
	long  = compressed sensing
}
\DeclareAcronym{rip}{
short = RIP,
long  = restricted isometry property
}
\DeclareAcronym{iid}{
	short = i.i.d.,
	long  = independent and identically distributed
}
\DeclareAcronym{rfid}{
	short = RFID,
	long  = radio-frequency identification
}
\DeclareAcronym{bp}{
	short = BP,
	long  = belief propagation
}
\DeclareAcronym{lasso}{
	short = lasso,
	long  = least absolute shrinkage and selection operator
}
\DeclareAcronym{amp}{
	short = AMP,
	long  = approximate message passing
}
\DeclareAcronym{bamp}{
	short = BAMP,
	long  = Bayesian approximate message passing
}
\DeclareAcronym{pdf}{
	short = pdf,
	long  = probability density function
}
\DeclareAcronym{mmse}{
	short = MMSE,
	long  = minimum mean squared error
}
\DeclareAcronym{mmv}{
	short = MMV,
	long  = multiple measurement vector
}
\DeclareAcronym{dcs}{
	short = DCS,
	long  = distributed compressed sensing
}
\DeclareAcronym{mri}{
	short = MRI,
	long  = magnetic resonance imaging
}
\DeclareAcronym{mse}{
	short = MSE,
	long  = mean squared error
}
\DeclareAcronym{map}{
	short = MAP,
	long  = maximum a posteriori
}
\DeclareAcronym{em}{
	short = EM,
	long  = expectation-maximization
}
\DeclareAcronym{bg}{
	short = BG,
	long  = Bernoulli-Gauss
}
\DeclareAcronym{se}{
	short = SE,
	long  = state evolution
}
\DeclareAcronym{ptc}{
	short = PTC,
	long  = phase transition curves
}
\DeclareAcronym{nmse}{
	short = NMSE,
	long  = normalized mean squared error
}
\DeclareAcronym{gevd}{
	short = GEVD,
	long  = generalized eigenvalue decomposition
}
\DeclareAcronym{vbamp}{
	short = MMV/DCS-BAMP,
	long  = MMV/DCS-Bayesian approximate message passing
}
\DeclareAcronym{Vbamp}{
	short = BAMP,
	long  = Bayesian approximate message passing
}
\DeclareAcronym{snr}{
	short = SNR,
	long  = signal-to-noise ratio
}
\DeclareAcronym{pt}{
	short = PT,
	long  = phase transition
}
\DeclareAcronym{jsm-2}{
	short = JSM-2,
	long  = joint sparsity model 2
}
\newtheorem{theorem}{Theorem}

\newtheorem{proof}{Proof}
\newtheorem{lemma}[theorem]{Lemma}

\usepackage{booktabs}

\allowdisplaybreaks

\begin{document}

\title{Performance Analysis of Approximate Message Passing for Distributed Compressed Sensing}

\author{Gabor Hannak, 
	Alessandro Perelli, 
	Norbert Goertz, \IEEEmembership{Senior Member,~IEEE,}
	Gerald Matz, \IEEEmembership{Senior Member,~IEEE,}
	and Mike E. Davies, \IEEEmembership{Fellow,~IEEE}	
	\thanks{G.\ Hannak, N.\ Goertz, and G.\ Matz are with the Institute of Telecommunications, Vienna University of Technology, Vienna, Austria (e-mail:  ghannak@nt.tuwien.ac.at, ngoertz@nt.tuwien.ac.at, gmatz@nt.tuwien.ac.at)}
	\thanks{A.\ Perelli and M.\ E.\ Davies are with the Institute of Digital Communications, University of Edinburgh, Edinburgh, UK, (e-mail: m.davies@ed.ac.uk; a.perelli@ed.ac.uk)}
	\thanks{This work was funded in part by WWTF Grant ICT15-119, ERC Grant 694888, and EPSRC Grant EP/M008916/1; MD is also supported through a Royal Society Wolfson Research Merit Award.}
	}

% make the title 
\maketitle
% ############################################################################

% ### Abstract

\begin{abstract}
%\boldmath
\Ac{bamp} is an efficient method in compressed sensing that is nearly optimal in the \ac{mmse} sense. Multiple measurement vector (MMV)-\Ac{bamp} performs joint recovery of multiple vectors with identical support and accounts for correlations in the signal of interest and in the noise. In this paper, we show how to reduce the complexity of vector \ac{bamp} via a simple joint decorrelation (diagonalization) transform of the signal and noise vectors, which also facilitates the subsequent performance analysis. We prove that %\ac{Vbamp} and 
the corresponding \ac{se} is equivariant with respect to the joint decorrelation transform and preserves diagonality of the residual noise covariance for the \ac{bg} prior. We use these results to analyze the dynamics and the \ac{mse} performance of \ac{bamp} via the replica method, and thereby understand the impact of signal correlation and number of jointly sparse signals. Finally, we evaluate an application of MMV-BAMP for single-pixel imaging with correlated color channels and thereby explore the performance gain of joint recovery 
compared to conventional BAMP reconstruction as well as {group lasso}.
\end{abstract}

% UNCOMMENT THIS FOR IEEE
%\IEEEpeerreviewmaketitle
\acresetall
\acuse{mmse}
% ##########################################################################################

\section{Introduction}
\Ac{cs} is a signal processing technique aiming at recovering a high-dimensional sparse vector from a (noisy) system of linear equations \cite{donoho2006compressed,CandesTao}. Joint sparsity refers to multiple vectors having the same support set\footnote{The support set of a vector consists of the indices of the vector's nonzero entries.}, whose cardinality is typically much lower than the signal dimension. There are two prominent \ac{cs} scenarios \cite{cotter2005sparse,duarte2005joint} in the context of joint sparsity: (i) the \ac{mmv} problem, where the measurement matrices are identical, and (ii) the \ac{dcs} problem, where the measurement matrices are independent. Joint sparsity arises in a number of real-world scenarios, e.g., when multiple sensors or antennas observe the same signal corrupted by different channels and noise (e.g., \cite{cotter2005sparse,duarte2005joint}). 
A prime example is radio frequency identification where the observed vectors are the received signals at different antennas (of the same receiver) \cite{mayer2016exploiting}. Additionally, typical applications are magnetic resonance imaging \cite{liang2009parallel}, distributed networks \cite{wimalajeewa2014omp}, wireless communications \cite{mayer2016exploiting}, and direction of arrival estimation \cite{tzagkarakis2010multiple}. 

In this work, we investigate an \ac{amp} solution for joint sparse recovery when there is possible correlation between the signals (and the noise). We then evaluate this algorithm in the context of single-pixel color imaging \cite{duarte2008sp}. In particular, we show the potential of joint recovery that exploits the correlation between the red, green, blue (RGB) color intensity channels. 

\subsection{Related Work}
Several methods for jointly sparse recovery have been proposed in the literature 
\cite{cotter2005sparse,tropp2006algorithms1,tropp2006algorithms2,wipf2007empirical,davies2012rank,schniter2010turbo,wimalajeewa2014omp,ziniel2013efficient,mayer2015bayesian,zhao2015joint,lu2016independent,kim2011belief}. \Ac{amp} was introduced in \cite{donoho2009message,DonohoITWmessagepassingI,DonohoITWmessagepassingII} as a large system relaxation of loopy belief propagation to solve a random linear system with sparsity constraint. Scalar \ac{bamp}, its Bayesian version \cite{maleki2010approximate,montanari2012graphical}, uses the signal prior explicitly and is an efficient approximate \ac{mmse} estimator. The turbo \ac{bamp} methods in \cite{schniter2010turbo,ziniel2013efficient,mayer2015bayesian}, and their generalization in \cite{rangan2017} for clustered sparse signals, improve the recovery performance by exchanging extrinsic information about the current support estimate in each message passing iteration.
In \cite{zhao2015joint,lu2016independent,zhu2016performance}, joint sparsity is directly enforced by an appropriate vector estimator (denoiser) function for the \ac{bg} prior. 

The \ac{se} formalism developed in \cite{DonohoITWmessagepassingI, DonohoITWmessagepassingII, bayati2011dynamics}  analytically predicts the recovery performance of (B)\acs{amp} algorithms. \ac{se} was employed to analyze \ac{bamp} for joint sparsity with a vector estimator and to point out the difference between the \ac{dcs} and \ac{mmv} scenarios in \cite{lu2016independent}. Recent works rigorously prove the \ac{se} for non-separable non-linearities \cite{berthier2017} and a class of sliding-window denoisers \cite{ma2017} with Gaussian i.i.d. measurement matrices. Furthermore, the \ac{se} of the Vector AMP has been derived for a large class of right orthogonally invariant random sensing matrices \cite{rangan2017vamp}. (We highlight that the acronym Vector AMP should not to be confused with the vector-prior version of BAMP, considered in this paper for the MMV/DCS problems.) 

In \cite{zhu2016performance}, the replica method (a statistical physics tool for large disordered systems) is used to calculate the \ac{mmse} of the \ac{cs} measurement (note that \cite{zhu2016performance} refers to \ac{mmv} and \ac{dcs} as MMV-2 and MMV-1, respectively).
The replica trick non-rigorously simplifies the high-dimensional integral for the \ac{mmse} of the Bayesian estimator of the \ac{cs} channel, thereby leading to the free energy as a function of the \ac{mse}. The local maxima in the free energy function correspond to stable fixed points of \ac{bp} and \ac{bamp} and thus predict the expected \ac{mse} of \ac{bamp}. The replica analysis in \cite{zhu2016performance} is performed for the \ac{bg} signal prior with uncorrelated isotropic unitary signal and uncorrelated isotropic Gaussian noise distribution, i.e., with a single noise parameter.

\subsection{Contributions}
We consider the vector-prior \ac{bamp} algorithm for the \ac{dcs} and \ac{mmv} problems, which uses an appropriate vector \ac{mmse} estimator function and Onsager correction term to exploit joint sparsity structure, the signal distribution, and the noise covariance. We provide an analytical performance prediction for the \ac{bamp} algorithm with a \ac{bg} signal prior with arbitrary signal and noise correlation by (i) incorporating a linear joint decorrelation of the measurements, (ii) showing the equivariance of \ac{Vbamp} w.r.t.\ invertible linear transformations, (iii) extending the replica analysis from \cite{zhu2016performance} to arbitrary diagonal noise covariance matrices.

In particular, the joint decorrelation yields a simpler equivalent measurement model with diagonal signal and noise covariance matrix (under mild conditions, one of the covariance matrices can be made the identity matrix). The simplified model naturally provides the measurement \acp{snr} of each signal vector and substantially reduces the complexity of the \ac{Vbamp} iterations. We further show that the \ac{Vbamp} algorithm is equivariant to invertible linear transformations, thus, it preserves its properties across iterations in the transformed domain and delivers a result equivalent to that obtained with the original measurements {and covariance parameters}. For the widely used \ac{bg} prior, we prove that the \ac{Vbamp} iterations (and the corresponding \ac{se}) preserve the diagonal structure of the (effective) noise covariance, thus implying that a $B$-dimensional state (instead of $B(B+1)/2$ dimensions) is sufficient and that every \ac{mmv} problem can be transformed into an equivalent \ac{dcs} problem. Finally, we extend the replica analysis in \cite{zhu2016performance} to the case of anisotropic noise (i.e., $B$ noise parameters instead of just $1$).
The replica analysis yields the $B$ measurement-wise \acp{mse} of the \ac{Vbamp} estimate in its {fixed} points.
{We use both real-world and synthetic images to compare MMV-BAMP to state-of-the-art scalar recovery algorithms and to joint sparsity-aware algorithms in the context of single-pixel color imaging.}

\subsection{Outline}
The remainder of this paper is organized as follows. In Section \ref{sec:vbamp}, we discuss the \ac{Vbamp} algorithm, the estimator function for the multivariate \ac{bg} signal prior, and the multivariate state evolution of \ac{Vbamp}. In Section \ref{sec:sec3}, the joint decorrelation of the signal and the noise vectors is investigated in the context of \ac{Vbamp} and state evolution; the multivariate \ac{bg} signal prior is studied as special case. In Section \ref{sec:replicanalysis}, we present the multivariate free energy formula for arbitrary diagonal noise covariance matrices (the details of the replica analysis are relegated to the appendix). Section \ref{sec:anisodyn} provides a qualitative discussion and open questions regarding the effects of signal correlation and the increasing number of jointly sparse vectors on the dynamics of \ac{Vbamp}. Section \ref{sec:sin_pix} evaluates the MMV-BAMP algorithm on a simplified single pixel imaging problem, highlighting the benefits of exploiting signal correlation across channels. We close with conclusions in Section \ref{sec:concl}.

\subsection{Notation}\label{subs:not}
Uppercase (lowercase) boldface letters denote matrices (vectors), and serif letters denote random quantities. For a matrix $\Av$ (vector $\av$), $\Av_{i}$ ($a_i$) denotes its $i$th row ($i$th entry) and $\av_{i}$ its $i$th column.
The all zero matrix and the identity matrix of dimension $M\times N$ are denoted by $\mbf{0}_{M\times N}$ and $\Iv_{M\times N}$, respectively (we omit the subscript if the dimensions are clear from the context). The Dirac delta (generalized) function is $\delta(\xv)$. The normal distribution with mean $\boldsymbol\mu$ and covariance matrix $\boldsymbol{\Sigma}$ is denoted by $\mcal{N}(\boldsymbol\mu,\boldsymbol{\Sigma})$ and $\mcal{N}(\mbf{x};\boldsymbol\mu,\boldsymbol{\Sigma})$ denotes the value of this normal \ac{pdf} at $\mbf{x}$. The outer product of a column vector $\xv$ with itself is denoted by $\boldsymbol{\langle} \xv \boldsymbol{\rangle} = \xv \xv^T$. For a vector $\xv=(x_1,\ldots,x_B)^T$, $\diag(\xv)$ and $\diag(x_1,\ldots,x_B)$ denote the diagonal matrix whose $i$th diagonal element equals $x_i$. For a matrix $\Xv$, $D(\Xv)$ is the diagonal matrix whose diagonal is identical to that of $\Xv$, i.e., $D(\cdot)$ is the orthogonal projection that zeros the off-diagonal elements. The Kronecker product of two matrices is denoted by $\otimes$.

% ##################################################################################################

\section{BAMP with Vector Denoiser}
\label{sec:vbamp}

\subsection{Measurement Model}

We consider the measurement model
\begin{equation}
 \yv(b) = \Av(b)\xv(b) + \wv(b)\,, \label{eq:matrixmeasurement}
\end{equation}
with $\yv(b)\in\mbb{R}^M$, $\xv(b)\in\mbb{R}^N$, $\wv(b)\in\mbb{R}^M$, and $\Av(b)\in\mbb{R}^{M\times N}$, for $b=1,\ldots,B$.
We denote the measurement rate by $R = {M}/{N}$.
We assume that the measurement matrices $\Av(b)$ are realizations of Gaussian or Rademacher random matrices \cite{foucart2013mathematical} with normalized columns.
If the measurement matrices $\Av(b)$ are identical (i.e., $\Av(b) = \Av$, $b=1,\dots,B$) we have an \ac{mmv} scenario; if they are mutually independent then we have a \ac{dcs} scenario.
We define the length-$B$ column vectors
\begin{align}
 \xvec_n &= (x_n(1),\ldots,x_n(B))^T, \nonumber\\ % \quad \forall n \,,\\
 \yvec_m &= (y_m(1),\ldots,y_m(B))^T, \\ % \quad \forall m \,,\\
 \wvec_m &= (w_m(1),\ldots,w_m(B))^T  \nonumber % \quad \forall m\,,
\end{align}
(similar notation will be used throughout the paper).
Joint sparsity (cf.\ JSM-2 in \cite{duarte2005joint}) with sparsity (or nonzero probability) $\eps$ requires that $\xvec_n=\zerov$ with probability $1-\eps$ and $\xvec_n\ne\zerov$ % or all its components are nonzero 
with probability $\eps$.
In this work, we focus on signals with multivariate \ac{bg} \ac{pdf}, i.e., 
\begin{equation}
f_{\xvecs_n}(\xvec_n) = f_{\xvecs}(\xvec_n) =
(1-\eps) \,
\delta(\xvec_n)
+ \eps\, \mcal{N}(\xvec_n;\zerov,\Sx) , \label{eq:signal_prior}
\end{equation}
\ac{iid} over $n$; here, $\Sx$ is the covariance matrix of $\xvec_n$ given that it is non-zero vector.
The additive noise in \eqref{eq:matrixmeasurement} is assumed to be \ac{iid}\ Gaussian over $m$ with zero mean and covariance $\Sw$,
\begin{equation}
	\wvecs_m \sim \mcal{N}(\zerov, \Sw) \,.
\end{equation}

\subsection{Vector-prior BAMP for MMV/DCS} \label{subsec:vbamp}

\begin{algorithm}[t]
	\caption{\acs{Vbamp} for MMV/DCS}
	\label{alg:vbamp}
	\begin{algorithmic}[1]
		\State {\bf input:} $\yv(b)$, $\Av(b)$, $\Sx$, $\vareps$, $t_{\max}$, $\vareps_{\tol}$ 
		\State $t=0$, $\xvechat_n^t=\zerov_{B\times 1}$, $\rvec_m^t=\yvec_m$, $\forall m, n$
		\Do
		\State $t \leftarrow t+1$ 
		\State $\uv^{t-1}(b)= \xvhat^{t-1}(b) + \Av(b)^T \rv^{t-1}(b)$, $\forall b$ 
		\State $\Svv^{t-1} = \begin{cases} {\Srr^{t-1}} %\cov\!\lef\{ \rvec_m^{t-1} \rig\} 
        &\mbox{ for MMV} \\ D\big( {\Srr^{t-1}} %\cov\!\lef\{ \rvec_m^{t-1} \rig\}
        \big) &\mbox{ for DCS} \end{cases}$ %{\color{gray} \Comment{Estimate effective noise variance}}
		\State $\xvechat_n^t = F(\uvec_n^{t-1} ; \Svv^{t-1})$, $\forall n$  
		\State $\rvec_m^{t} = \yvec_m - \big(\Av(1) \xvhat^t(1),\ldots,\Av(B) \xvhat^t(B)\big)_m \hspace*{15mm}$ 
		\hspace*{10mm} $\phantom{=}+  \frac{1}{M} \sum_{n=1}^{N} F'(\uvec_n^{t-1} ; \Svv^{t-1}) \rvec_m^t$, $\forall m$
		\doWhile{$\sum_{b=1}^{B}\left\|\xvhat^{t}(b)\!-\!\xvhat^{t-1}(b) \right\|^2_2  \!> \! \vareps_{\tol} \sum_{b=1}^{B}\left\|\xvhat^{t-1}(b)\right\|^2_2$ {\bf and} $t < t_{\max}$}  %{\color{gray} \Comment{Check stopping criteria}}
		\State
		\Return $\xvhat(b)=\xvhat^t(b)$, 
		$\forall b$ %{\color{gray},  \Comment{Recovered vectors}}
	\end{algorithmic}
\end{algorithm}

The \ac{Vbamp} method for joint sparse recovery of $\xv(b)$, $b=1,\dots,B$,
\cite{kim2011belief,zhao2015joint} is summarized in Algorithm \ref{alg:vbamp}
(superscript $t$ indicates the iteration index).
Note that scalar \ac{Vbamp} (i.e., when $B=1$) is a special case of Algorithm \ref{alg:vbamp} where \ac{mmv} and \ac{dcs} are equivalent.
The vector-prior BAMP follows similar steps as ordinary scalar \ac{bamp} \cite{donoho2009message,maleki2010approximate,montanari2012graphical,DonohoITWmessagepassingI,DonohoITWmessagepassingII,bayati2011dynamics}. According to the decoupling principle \cite{montanari2012graphical}, which holds in the asymptotic regime where $M,N\rightarrow \infty$ while $\frac{M}{N} = R$, the \ac{Vbamp} algorithm decouples the \ac{cs} measurements \equref{eq:matrixmeasurement} according to
\begin{equation} % \nonumber
\uvec_n^t=\xvec_n + \vvec_n^t ,
\end{equation}
where the effective noise vector is distributed as ${\vvecs_n^t \sim \mcal{N}(\zerov, \Svv^t)}$. 
The effective noise covariance is estimated via the empirical covariance {$\Srr^{t}=\cov\!\lef\{ \rvec_m^{t-1} \rig\}$} from vectors $\rvec_m^{t-1}$ in line 6 of Algorithm \ref{alg:vbamp}. It has been shown in \cite{lu2016independent} that in the \ac{dcs} scenario only the diagonal entries of the covariance matrix are retained due to the mixing effected by the $B$ {mutually} independent measurement matrices. In the following, we will simplify notation by
occasionally dropping the indices $t$ and $n$.

The vector denoiser in \ac{Vbamp} (line 7 of Algorithm \ref{alg:vbamp}) amounts to a vector \ac{mmse} estimator of $\xvec_n$ given the decoupled measurements $\uvec_n$.
Using Bayes' theorem, the denoiser can be written as:
\begin{equation}
\begin{aligned}
F(\uvec; \Svv) &= \Exp_{\xvecs} \lef\{\xvecs \mid \uvecs = \uvec ; \Svv \rig\}\\
&= \frac{\int_{\mathbb{R}^{B}} \zvec \,\mathcal{N}(\uvec;\zvec,\Svv) f_{\xvec}(\zvec) \,d \zvec}{\int_{{\mathbb{R}^{B}}} \mathcal{N}(\uvec;\zvec,\Svv) f_{\xvec}(\zvec) \,d \zvec} \,,
\end{aligned}
\label{eq:F}
\end{equation}
{where the covariance of the effective noise is $\Svv = \Srr$ (\ac{mmv}) or $\Svv = D(\Srr)$ (\ac{dcs})}.
For the multivariate \ac{bg} prior \equref{eq:signal_prior}, the vector denoiser becomes
\begin{align}
F(\uvec; \Svv) = \Wv\uvec \qquad\text{with }\;
\Wv = \frac{F_N(\uvec; \Svv)}{F_D(\uvec; \Svv)}\Sx \Su^{-1} .
\label{eq:2_Fvec_bernoulli_gaussian}
\end{align}
Here, $\Su = \Sx + \Svv$ and
\begin{align}\label{eq:BGprior}
F_N(\uvec; \Svv) & = \eps\, \mcal{N}(\uvec;\zerov,\Su),\\
F_D(\uvec; \Svv) & = (1-\eps)\, \mcal{N}(\uvec;\zerov,\Svv) + \eps\, \mcal{N}(\uvec;\zerov,\Su)\nonumber
\end{align}
The denoiser \equref{eq:2_Fvec_bernoulli_gaussian} consists of a multivariate Gaussian Wiener estimator followed by a joint shrinkage operation.

The \ac{Vbamp} residual is computed in line 8 of Algorithm \ref{alg:vbamp}.
As in the original \ac{amp} derivation \cite{maleki2010approximate}, the Onsager correction term for the residual $\yvec_m - \big(\Av(1) \xvhat(1),\ldots,\Av(B) \xvhat(B)\big)_m$ is computed via the derivative of the estimator. 
In the asymptotic regime, the Onsager term 
\begin{equation}\label{eq:onsager}
	\frac{1}{M} \sum_{n=1}^{N} F'(\uvec_n ; \Svv) \,\rvec_m 
\end{equation}
renders the decoupled measurement vectors $\uvec_n$ Gaussian with mean $\xvec_n$ and covariance $\Svv$  \cite{kim2011belief,zhu2016performance}.
Here, the Jacobian matrix $F'(\uvec ; \Svv) = {dF(\uvec; \Svv)}/{d\uvec^T}  $ of the estimator $F(\uvec ; \Svv)$ is given by
\begin{equation}
F'(\uvec; \Svv) % & = \frac{dF(\uvec; \Svv)}{d\uvec^T}  \\
				\! =\!
\Wv \!-\! \left(1 \!-\! \frac{F_N(\uvec; \Svv)}{F_D(\uvec; \Svv)}\right) \Wv \uvec_n \uvec_n^T (\Su^{-1}-\Svv^{-1})\,.
\end{equation}

The algorithm runs until the relative change in the estimated signal is below a certain threshold $\vareps_{\tol}$ or the maximum number of iterations $t_{\max}$ is reached.
Compared to scalar \ac{bamp}, the vector \ac{Vbamp} algorithm involves the following crucial modifications:
\begin{itemize}
\item a multivariate prior (possibly with joint sparsity structure and correlation);
\item the estimator acts on vectors rather than scalars \equref{eq:F} and both correlated signal and correlated additive noise are taken into consideration (more precisely, the full signal and noise vector \ac{pdf} is taken into account);
\item an % matrix-valued 
	  Onsager term obtained as the sum of Jacobian matrices (cf.\ \equref{eq:onsager}).
\end{itemize}

\subsection{State Evolution}

\ac{se} was originally proposed in \cite{donoho2009message} for scalar (B)AMP and extended to the \ac{mmv} and \ac{dcs} scenarios (e.g., in \cite{lu2016independent}); it allows to characterize analytically the expected behavior of \ac{Vbamp}
(note that the Onsager term in \cite{lu2016independent} is flawed 
% defined incorrectly. Nonetheless, the presentation of 
even though the multivariate \ac{se} is correct).
In particular, the \ac{se} equation predicts the evolution of the effective noise covariance (the state)
for any signal prior {$f_{\xvecs}(\xvec_n)$} as
%  across successive iterations:
\begin{align}
\Svv^{t+1} 
%& \nonumber \\
%&\hspace*{-10mm	}
= \begin{cases}
\Sw + \frac{1}{R} \Exp_{\xvecs,\vvecs} \lef\{ \boldsymbol{\langle} \ev(\xvecs,\vvecs) \boldsymbol{\rangle} \rig\} & \mbox{for MMV},\\
%\Sw + \frac{1}{R} \Exp_{\xvecs,\vvecs} \lef\{ \boldsymbol{\langle} G(\xvecs+\vvecs;\Svv^{t}) - \xvecs \boldsymbol{\rangle} \rig\} & \mbox{for MMV},\\
D\!\left(\Sw + \frac{1}{R} \Exp_{\xvecs,\vvecs} \lef\{ \boldsymbol{\langle} \ev(\xvecs,\vvecs)\boldsymbol{\rangle}\rig\}\right) & \mbox{for DCS} ,
%D\!\left(\Sw + \frac{1}{R} \Exp_{\xvecs,\vvecs} \lef\{ \boldsymbol{\langle} G(\xvecs+\vvecs;\Svv^{t}) - \xvecs\boldsymbol{\rangle}\rig\}\right) & \mbox{for DCS} ,
\end{cases}  \label{eq:stateevolution}
%& = \Sigma_{\wvec} + \frac{N}{M} \frac{dG}{d(\xvec+\vvec^t)} \Sigma^t_{\vvec}
\end{align}
where $\ev(\xvecs,\vvecs)=F(\xvecs+\vvecs;\Svv^{t}) - \xvecs$ is the error achieved by the
\ac{mmse} estimator $F(\uvec;\Svv^t)$ and $\vvecs \sim \mcal{N}(\zerov,\Svv^t)$.
The state in the \ac{mmv} scenario is in general $B(B+1)/2$ dimensional (since the covariance matrix is symmetric).
From \equref{eq:stateevolution}, the \ac{mse} prediction directly follows as
\begin{align*}
\cov \{\uvec^t_n - \xvec_n\} &= \Svv^t , \\
%\cov \{\xvechat^t_n - \xvec_n\} &= R (\Svv^t - \Sigma_{\wvecs}) \\
\widehat{\mathrm{MSE}}{}^t(b) &= R\,(\Svv^t - \Sw)_{b,b} ,
\end{align*}
with the \ac{mse} per channel being defined as 
\begin{equation*}
\mathrm{MSE}^t(b) = \frac{1}{N}\|\xvhat^t(b) - \xv(b)\|_2^2 .
\end{equation*}

% ##################################################################################################

% ### GEVD
\section{Diagonalized Vector-prior BAMP} \label{sec:sec3}

\subsection{Joint Diagonalization for MMV} \label{sec:gevd}

\begin{algorithm}[!t]
	%\small
	\caption{joint diagonalization transformation}
	\label{alg:gevd}
	\begin{algorithmic}[1]
		\State Given $\Sx,\Sw$
		\State \hspace*{5mm} find $\Pv$ such that $\Pv\Pv^T=\Sw$
		\State \hspace*{5mm} $\Gv = \Pv^{-1} \Sx \Pv^{-T}$
		\State \hspace*{5mm} find eigendecomposition $\Qv \boldsymbol{\Lambda} \Qv^T=\Gv$
		\State \hspace*{5mm} $\Tv = \boldsymbol{\Lambda}^{-1/2}\Qv^T\Pv^{-1}$
	\end{algorithmic}
\end{algorithm}

The \ac{Vbamp} algorithm in Section \ref{subsec:vbamp} can deal with arbitrary signal and noise correlations $\Sx$ and $\Sw$, which in general results in a nondiagonal $\Svv^t$ in the \ac{mmv} scenario.
In the decoupled measurements $\uvec=\xvec + \vvec$, it means that there are $\mcal{O}(B^2)$ \ac{snr} relations and $B(B+1)/2$ states: each $\xs_n(b)$ correlates with all $\xs_n(b')$, $b'\in\{1,\ldots,B\}\setminus\{b\}$, and it is influenced simultaneously by all effective noise components $\vs(b')$, $b'\in\{1,\ldots,B\}$.

Under the assumption that the covariance matrices $\Sx$ and $\Sw$ are full rank and using the fact that covariance matrices are symmetric and positive definite and \cite[Thm.~7.6.1.]{horn2012matrix}, there exists a nonsingular (but generally non-orthogonal) matrix $\Tv$ that simultaneously diagonalizes the covariance matrices of the signal $\xvec$ and the noise $\wvec$.
The computation of $\Tv$ is described in Algorithm \ref{alg:gevd}. 
In the transformed model
\begin{align}
\tilde{\yvec}_m = \Tv \yvec_m , \quad \tilde{\xvec}_n &= \Tv \xvec_n , \quad \tilde{\wvec}_m = \Tv \wvec_m \label{eq:transformedmodel2} ,
\end{align}
we thus have 
\begin{align*}
\Sxt & = \Tv \Sx \Tv^T = \Iv_{B\times B} , \\
\Swt & = \Tv \Sw \Tv^T = \boldsymbol{\Lambda}^{-1}=\eps\diag \lef(\frac{1}{\snr(1)}, \ldots, \frac{1}{\snr(B)}\rig) .
%\cov\{\Tv\xvecs_n \} &=  \eps\,\Tv \Sx \Tv^T  % \eps\,\Sxt 
%                        = \eps\,\Iv_{B\times B} \,,\\
%\cov\{\Tv\wvecs_m \} &= \Tv \Sw \Tv^T = \Swt = \diag \lef(\frac{1}{\snr(1)}, \ldots, \frac{1}{\snr(B)}\rig) \,,
\end{align*}
Here, the per-channel SNRs are defined as
\begin{equation*}
	\mathrm{SNR}(b) = \frac{\Exp_{\xvs}\! \left\{\|\Av(b)\tilde{\xvs}(b)\|^2_2\right\}}{\Exp_{\wvs} \!\left\{\|\tilde{\wvs}(b)\|^2_2\right\}}  = \eps\,\boldsymbol{\Lambda}_{b,b}.
% 	\frac{\eps (\Sxt)_{b,b}}{(\Swt)_{b,b}}\,.
\end{equation*}
Note that the decorrelation can be applied also in the \ac{dcs} scenario, given that only the noise covariance $\Sw$ is nondiagonal and the signal covariance $\Sx$ is diagonal.
We emphasize that in case \ac{Vbamp} operates on the transformed measurements, 
% \equref{eq:transformedmodel1}, in general 
the change in the prior distribution has to be accounted for in a nontrivial manner.
That is, the \ac{mmse} estimator \equref{eq:F} and its derivative will have a different form.
Consider the \ac{se} equation \equref{eq:stateevolution} that describes the expected evolution of the effective noise covariance over the \ac{Vbamp} iterations.
In the \ac{mmv} scenario, % apart from some special cases, 
even if $\Sw$ and $\Svv^t$ are diagonal, $\Svv^{t+1}$ in general will not be diagonal because the estimator $F(\uvec^t_n,\Svv^t)$ operates on the overall vector $\uvec^t_n$ ({nonetheless} the diagonalization described in Algorithm \ref{alg:gevd} could be performed repeatedly in each iteration).
However, we will see shortly that in the particular case of the \ac{bg} prior this is no longer the case.
A direct calculation reveals that
\begin{equation*}
\cov\{ \yvecs_m \} =
\begin{cases}
\Sw + \frac{1}{R}\cov\{\xvecs_n\}  &\mbox{for MMV}, \\
\Sw + \frac{1}{R} D\lef( \cov\{\xvecs_n\} \rig) &\mbox{for DCS}.
\end{cases} 
\end{equation*}
Thus, if we know either the noise or signal covariance then the other can be estimated directly through the measurement covariance $\cov\{\yvecs_m \}$. Alternatively, when both covariances are unknown and the signal is drawn from a BG prior we can use the \ac{em} AMP approach introduced in \cite{SchniterEMAMP} to estimate both sets of parameters within the iterations.
 
\subsection{Equivariance of \ac{Vbamp} for MMV}
We next establish the fact that for MMV both \ac{Vbamp} and its \ac{se} are equivariant w.r.t.\ invertible linear transformations of the input. The proof of this result is provided in Appendix \ref{app:vbamp_equivariance}.

\vspace*{2mm}
\begin{theorem}\label{thm:equivar}
	Algorithm \ref{alg:vbamp} for \ac{mmv} and its \ac{se} are equivariant w.r.t.~invertible linear transformations. 
	% as defined in \equref{eq:transformedmodel1} and \equref{eq:transformedmodel2}.
	Denote one \ac{Vbamp} iteration by $(\xvechat_n^{t+1}, \rvec_m^{t+1}, \Svv^{t+1})=\text{V}(\yvec_m, \xvechat_n^t, \rvec_m^t, \Svv^t)$. For any nonsingular $\Tv$, we have for all $m$ and $n$
\[
\text{V}(\Tv\yvec_m, \Tv\xvechat_n^t, \Tv\rvec_m^t, \Tv\Svv^t\Tv^T) = (\Tv\xvechat_n^{t+1}, \Tv\rvec_m^{t+1}, \Tv\Svv^{t+1}\Tv^T).
\]
Furthermore, the SE equation \eqref{eq:stateevolution} translates to the transformed domain as
\begin{align}
&\Tv\Svv^{t+1}\Tv^T  =  \Tv\Sw\Tv^T \nonumber\\
& \hspace*{10mm}+ \frac{1}{R}  \Exp_{\xvecs,\vvecs} \lef\{ \boldsymbol{\langle} F(\Tv(\xvecs+\vvecs);\Tv{\Svv}^{t}\Tv^T) - \Tv\xvecs\boldsymbol{\rangle}  \rig\} .\label{eq:transformedSE1}
\end{align}
\end{theorem}
\vspace*{1mm}

Note that \equref{eq:transformedSE1} holds for any signal prior in the Bayesian setting, i.e., when the estimator is the \ac{mmse} estimator.
Assume that \ac{Vbamp} converges to $\xvechat_n$ with inputs $\yvec_n$, $\Sx$, and $\Sw$;
then, Theorem \ref{thm:equivar} implies that \ac{Vbamp} with inputs $\Tv\yvec_n$, $\Tv\Sx\Tv^T$, and $\Tv\Sw\Tv^T$ 
converges to the solution $\Tv\xvechat_n$. 

\subsection{Bernoulli-Gauss Prior} \label{sec:bg}

For the \ac{bg} prior, after applying the transformation $\Tv$, the equivalent measurement model becomes
\begin{equation}
\tilde{\yv}(b) = \Av(b) \tilde{\xv}(b) +  \tilde{\wv}(b) \,, \quad \forall b \label{eq:equivalentmodel}
\end{equation}
with signal and noise \acp{pdf}
\begin{align}
f_{\tilde{\xvecs}}(\tilde{\xvec}_n) &=
(1-\eps) \,\delta(\tilde{\xvec}_n)
+ \eps\, \mcal{N}(\tilde{\xvec}_n;\zerov,\Iv) , \label{eq:diagonalbgpdf} \\
f_{\tilde{\wvecs}}(\tilde{\wvec}_m) &=
\mcal{N}(\tilde{\wvec}_m;\zerov,\boldsymbol{\Lambda}^{-1}) \label{eq:diagonalnoisepdf} .
\end{align}
That is, we retain a \ac{bg} prior in the transformed domain, only with uncorrelated components. 
This is a distinctive feature of the \ac{bg} prior and in general doesn't hold for other types of distributions.

In Appendix \ref{app:bg_diagonality} we demonstrate that 
for the decorrelated model \equref{eq:equivalentmodel} with \ac{bg} prior \equref{eq:diagonalbgpdf}--\equref{eq:diagonalnoisepdf}, the \ac{Vbamp} iterations under the MMV model preserve the diagonal structure  of $\Svvt^t$.
It follows that for \ac{cs} measurements with multivariate \ac{bg} signal prior, the decorrelation transformation has to be done only once before recovery; % and, if $B$ is not large, 
determining $\Tv$ itself is of negligible computational effort unless $B$ is very large.
These observations have the following implications:
\begin{itemize}
	\item The computation of \equref{eq:2_Fvec_bernoulli_gaussian} and \equref{eq:onsager} is significantly simplified, 
	leading to complexity reductions by a factor of $B$.
\item The dimension of the \ac{se} equations is $B$ instead of $B(B+1)/2$.
	In other words, $B(B+1)/2$ effective noise covariance parameters in $\Svv$ are reduced to $B$ effective noise variances, which % in turn carry naturally 
	explicitly characterize the \ac{mse} for each signal vector estimate as
	\begin{equation*}
%	\diag \lef(\mathrm{MSE}^t(1),\ldots,\mathrm{MSE}^t(B) \rig) = D\big(R (\Svvt^{t} - \Swt)\big) \,. 
	\widehat{\mse}{}^t(b) = R\, (\Svvt^{t} - \Swt)_{b,b}. %  \,,\quad \forall b
	\end{equation*}
	\item Every \ac{mmv} problem has an equivalent \ac{dcs} problem with possibly rescaled \acp{snr}.
	Furthermore, the analysis of \ac{dcs} also covers that of \ac{mmv}.
\end{itemize}

% ##################################################################################

\begin{figure}[t]
\centering  
\includegraphics[width=0.9\linewidth]{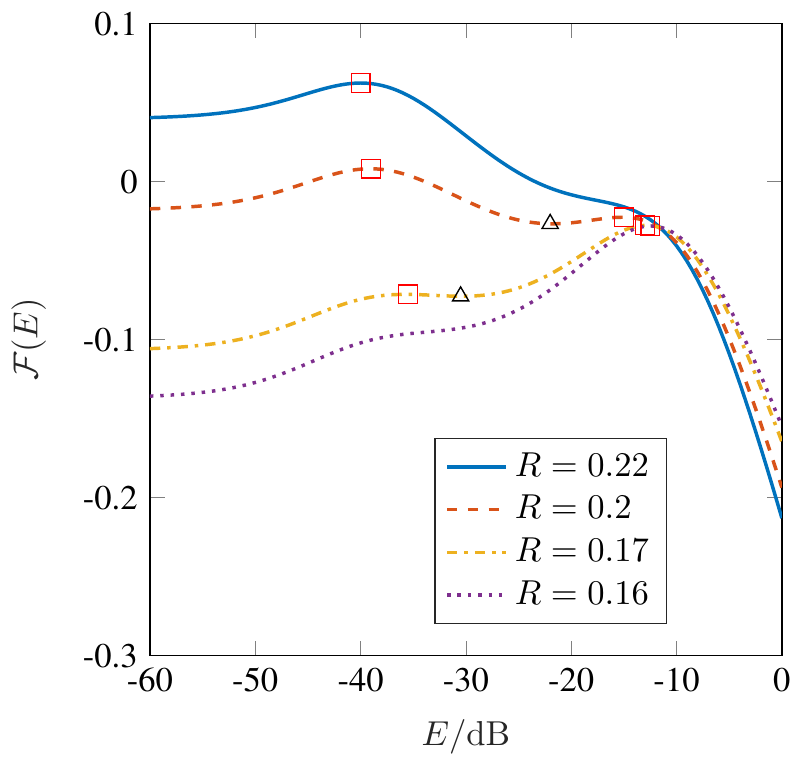}
\vspace*{2mm}
	\caption{{Free energy function at different rates $R$ for $B=1$, $\sw^2=-35\dB$, and sparsity $\eps=0.1$. Red squares and black triangles indicate local maxima and minima, respectively.}} \label{fig:freeenergycurves_noisy}
\end{figure}

\section{Replica Analysis}\label{sec:replicanalysis}

In \cite{zhu2016performance}, the replica method was used to determine the \ac{mse} performance of \ac{Vbamp} for the measurement \equref{eq:matrixmeasurement} and the \ac{bg} prior \equref{eq:signal_prior}, assuming $\Sx = \Iv$ and isotropic uncorrelated noise, i.e., $\Sw=\sigma_{\ws}^2 \Iv$.
In this special case \ac{mmv} and \ac{dcs} (referred to as {MMV-2} and {MMV-1}, respectively, in \cite{zhu2016performance}) are equivalent.
The analysis is quite sophisticated and the generalization to arbitrary signal and noise correlations seems infeasible.
However, due to the joint diagonalization approach from Section \ref{sec:sec3}, it suffices to extend the replica analysis to the case with $\Sx = \Iv$ and $\Sw=\diag( \sw^2(1),\ldots,\sw^2(B) )$.
In particular, the replica method is capable of predicting the fixed points of \ac{Vbamp} in the asymptotic regime ($N,M\rightarrow\infty$, $R=M/N=\mathrm{const.}$), as a function of the set of $B$ \acp{mse}\cite{krzakala2012probabilistic,krzakala2012statistical}.
We note that rigorous equivalence between the replica method and \ac{se} is not always guaranteed and requires additional technicalities \cite{reeves2016replica}.
Assuming $\Sx = \Iv$ and $\Sw = \diag\big(\sw^2(1),\ldots, \sw^2(B)\big)$, we compute in 
Appendix \ref{app:replica}, following the derivation in \cite{zhu2016performance}, the free energy  $\mcal{F}(\Evec)$
as a function of the \ac{mse} vector $\Evec = (E(1),\ldots,E(B))^T$ with $E(b)= \mathrm{MSE}(b)$, resulting in
\begin{equation}\label{eq:freeenergy}
\begin{split}
% & \mcal{F}(\Evec,\Sw) = 
& \mcal{F}(\Evec) = 
		(1-\eps)\,\zeta(\gamma) + \eps\,\zeta\Big(\frac{\gamma}{1+\gamma}\Big) \\
& {}-\frac{R}{2} \sum_{b=1}^{B}\lef(\log \frac{2\pi R}{\gamma(b)}  
		+ \gamma(b) \sw^2(b) 
		- \frac{1-\eps}{R} \frac{\gamma(b)}{1+\gamma(b)} 
		\rig) \!.
\end{split}
\end{equation}
In this expression we used 
\begin{align*}
\zeta(\eta) 
& = \int \log \left( \eps \prod_{b=1}^{B}(1+\gamma(b))^{-\frac12} \right. \\
&\qquad	+ \left. (1-\eps) \exp\!\bigg(\!-\frac12 \sum_{b=1}^{B} \eta(b)\,h^2(b)\bigg) \right)\mcal{D}\hv
\end{align*}
with
\begin{equation*}
	\gamma(b) = \frac{R}{E(b)+R\sw^2(b)};
\end{equation*}
furthermore, $\mcal{D}\hv = \mcal{N}(\hv;\zerov,\Iv)\,dh_1\dots dh_B$ denotes the multivariate standard Gaussian measure.

\begin{figure}[t]
\centering  
\includegraphics[width=0.9\linewidth]{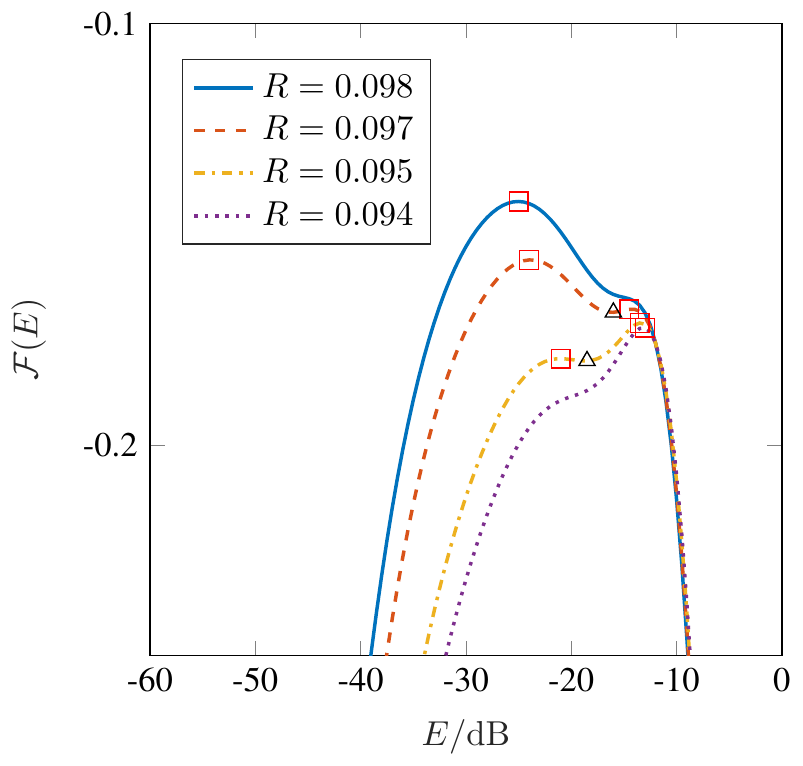}
\vspace*{2mm}
	\caption{{One-dimensional free energy function for the isotropic case with $B=10$ jointly sparse \ac{bg} vectors at rates around the phase transition rate ($\Sw = -35\dB \Iv_B$).}}\label{fig:freeenergy_B10}
\end{figure}

The stationary points of $\mcal{F}(\Evec)$ correspond to fixed points of belief propagation \cite{heskes2003stable}, and hence to those of \ac{Vbamp} in the asymptotic regime \cite{zhu2016performance}. Thus, we can determine the component-wise \ac{mse}s of \ac{Vbamp} by evaluating \equref{eq:freeenergy} and finding the largest components of $\Evec$ that correspond to a local maximum of $\mcal{F}(\Evec)$ \cite{yedidia2005constructing,yedidia2001bethe}. Note that for isotropic noise ($\sw^2(b) = \sw^2$ $\forall b$), the free energy in \equref{eq:freeenergy} simplifies to the result obtained in \cite{zhu2016performance} with one-dimensional argument $E=E(1)=\ldots=E(B)$. Replica curves for the isotropic case with $B=1$ and $B=10$ are shown in Figure~\ref{fig:freeenergycurves_noisy} and \ref{fig:freeenergy_B10} respectively. It is important to point out here that all the plots are the result of numerical integrations (and not Monte Carlo simulations). In the free energy function, local maxima correspond to stable fixed points and local minima to unstable fixed points, whereas the global maximum of $\mathcal{F}(E)$ corresponds to the \ac{mmse}. \ac{Vbamp} typically achieves the largest \ac{mse} associated with a local maximum.

\subsection{MMSE Gap}

In the \ac{cs} regime of small $\eps$ and nonzero noise variance, the \ac{mmse} estimate $\xvhat$ for a single measurement features a first order phase transition (PT) characterized by an abrupt change of the \ac{mse} at a certain rate $R_{\PT}$: for rates less than $R_{\PT}$, the \ac{mse} tends to be large, whereas for rates larger than $R_{\PT}$ the \ac{mse} tends to be small and plateaus to fixed nonzero value. This phenomenon can be seen in Figure \ref{fig:freeenergycurves_noisy}: for rates below $R\approx 0.16$, where the free energy has a single maximum at an \ac{mse} of about $-12\dB$ whereas for rates larger than $R\approx 0.17$  a second local maximum at \acp{mse} less then about $-37\dB$ appears. 

\begin{figure}[t]
\centering  
\includegraphics[width=0.9\linewidth]{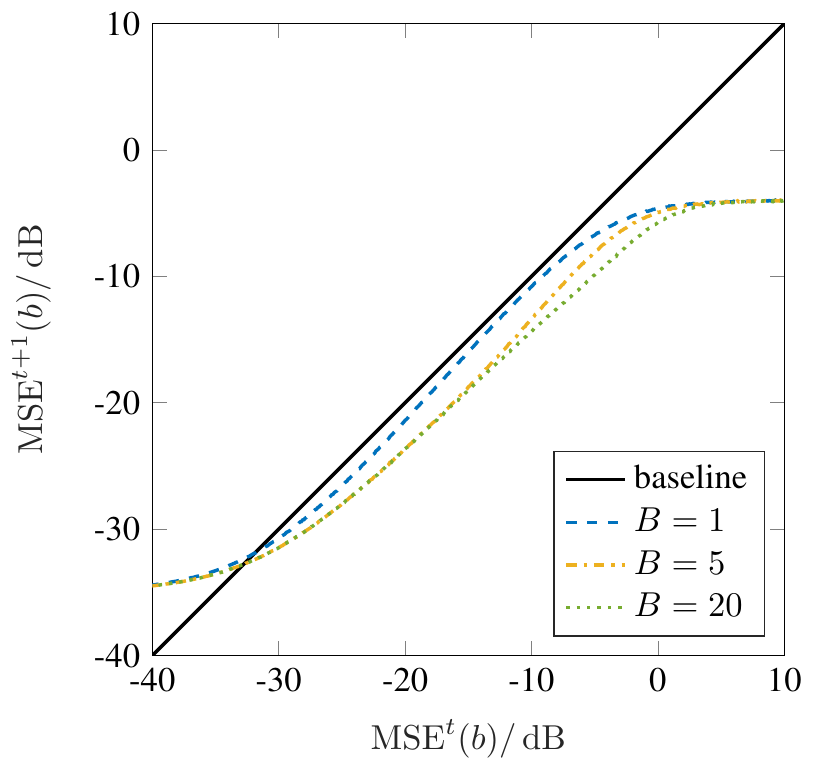}
\vspace*{2mm}
	\caption{Noisy \ac{se} curves for different number of jointly sparse \ac{bg} signals ($\eps=0.1$, $\Sx=\Iv_B$, $\Sw=-35\dB \Iv_B$, $R=0.25$).} \label{fig:SEcurves_B}
\end{figure}

A similarly abrupt phase transition does not appear to occur when the number of measurements $B$ is sufficiently large. Figure \ref{fig:SEcurves_B} shows the \ac{se} curves for various $B$ with $\eps=0.1$ and $\Sx=\Iv_B$. Observe that the ``bump'' in the \ac{se} curve for small $B$ and large \ac{mse}, which corresponds to the first 
fixed point, flattens out with increasing $B$. For large enough $B$ we observe that the \ac{se} curve ceases to exhibit a first order PT, so that the \ac{mse} changes smoothly with increasing rate $R$.

The same conclusion can be obtained by investigating the behavior of the free energy functions.  \ac{Vbamp} typically achieves the largest \ac{mse} which corresponds to a local maximum in the free energy, whereas the \ac{mse} at the global maximum of the free energy is the \ac{mmse}. As pointed out in \cite{zhu2016performance}, whenever the free energy function has a second local maximum at a larger \ac{mse} than the global maximum, \ac{Vbamp} is not Bayesian-optimal (i.e., does not reach the \ac{mmse}). For $B=1$, in Figure~\ref{fig:freeenergycurves_noisy}, a second local (non-global) maximum appears and thus BAMP is not MMSE optimal in the rate region $0.19<R<0.21$, while for $B=10$ with isotropic noise and sparsity $\epsilon=0.1$ it occurs at $R=0.097$ as shown in Figure \ref{fig:freeenergy_B10}. We speculate that the vanishing of the first order PT for sufficiently large $B$ may be a typical behaviour and something worthy of further investigation. 

While the possibility of no phase transition might appear surprising this relies on the presence of finite measurement noise. In such a setting there is no exact recovery PT. It would be interesting to understand what happens when the noise tends to zero, and see if comparisons could be drawn with PT results for the related problem of block sparse recovery \cite{taeb2013}, \cite{donoho2013accurate}. However, under this scenario it is not clear what would be the role of any anisotropy in the covariance matrices.

Finally, we emphasize that while our analysis here is asymptotic in the large system limit ($N,M,\eps N \rightarrow \infty$), it is non-asymptotic in the number of jointly sparse vectors $B$ which are assumed to be $\mcal{O}(1)$. This is in contrast to existing work \cite{blanchard2012recovery,kim2012compressive}, where results on the PT like phenomena were derived for the asymptotic case where $B\rightarrow\infty$ as $N\rightarrow \infty$.

\section{Anisotropic \ac{Vbamp} Dynamics}\label{sec:anisodyn}

We now consider the anisotropic scenaro.

\subsection{Correlated CS} \label{corrCS}

The matrix $\Tv$ from Algorithm \ref{alg:gevd} simultaneously decorrelates the signal and the noise.
While $\Tv\Sx\Tv^T=\Iv$, the transformed noise covariance $\Swt=\Tv\Sw\Tv^T$ depends on $\Sx$ and $\Sw$ in a nontrivial way
unless $\Sx$ and $\Sw$ commute. In this case, they have identical eigenvectors, i.e.,
$\Sx=\Qv\boldsymbol{\Lambda}_{\xvec}\Qv^T$ 
and $\Sw=\Qv\boldsymbol{\Lambda}_{\wvec}\Qv^T$, and we can show
\[
\Swt = \boldsymbol{\Lambda}_{\wvec}\boldsymbol{\Lambda}_{\xvec}^{-1}
= \diag\bigg( \frac{\lambda_{\wvec}(1)}{\lambda_{\xvec}(1)},\dots,\frac{\lambda_{\wvec}(B)}{\lambda_{\xvec}(B)} \!\!\bigg).
\]
Special cases of this situation occur when (i) either $\Sx$ or $\Sw$ is a scaled identity matrix 
and (ii) when both $\Sx$ and $\Sw$ are diagonal. The per-channel \acp{snr} are then obtained from $\Swt$ 
as $\snr(b)=\eps\lambda_{\xvec}(b)/\lambda_{\wvec}(b)$. While this result does not hold when
$\Sx$ and $\Sw$ do not commute, it is possible to derive the bounds
\[
\eps\,\frac{\min_k\{\lambda_{\xvecs}(k)\}}{\max_k\{\lambda_{\wvecs}(k)\}}
\leq \snr(b) \leq 
\eps\,\frac{\max_k\{\lambda_{\xvecs}(k)\}}{\min_k\{\lambda_{\wvecs}(k)\}}.
% \frac{\eps\min(\lambda_{\xvecs})}{\max(\lambda_{\wvecs})}.
\]

If a subset $\xv(b_1),\ldots,\xv(b_K)$ of the $B$ signal vectors is fully correlated, 
then $K-1$ of the \acp{snr} equal $0$.
Thus, the model is equivalent to one with $B-K+1$ (instead of $B$) measurements, but with different \acp{snr}.
The free energy function leads to the same conclusion: when taking the limits $\sw^2(b_1) = \ldots = \sw^2(b_{K-1}) \rightarrow \infty$ in the $B$-dimensional free energy function \equref{eq:freeenergy}, it can be seen that $\mcal{F}(\Evec)$ is independent of $E(b_1),\ldots,E(b_{K-1})$.
Therefore, the curvature of $\mcal{F}(\Evec)$ 
and hence the location of its stationary points
do not depend on those arguments, such that the $B$-dimensional free energy function effectively collapses into a $B-K+1$-dimensional function.

\begin{figure*}[t]
\hspace*{2mm}
\includegraphics[height=0.48\linewidth]{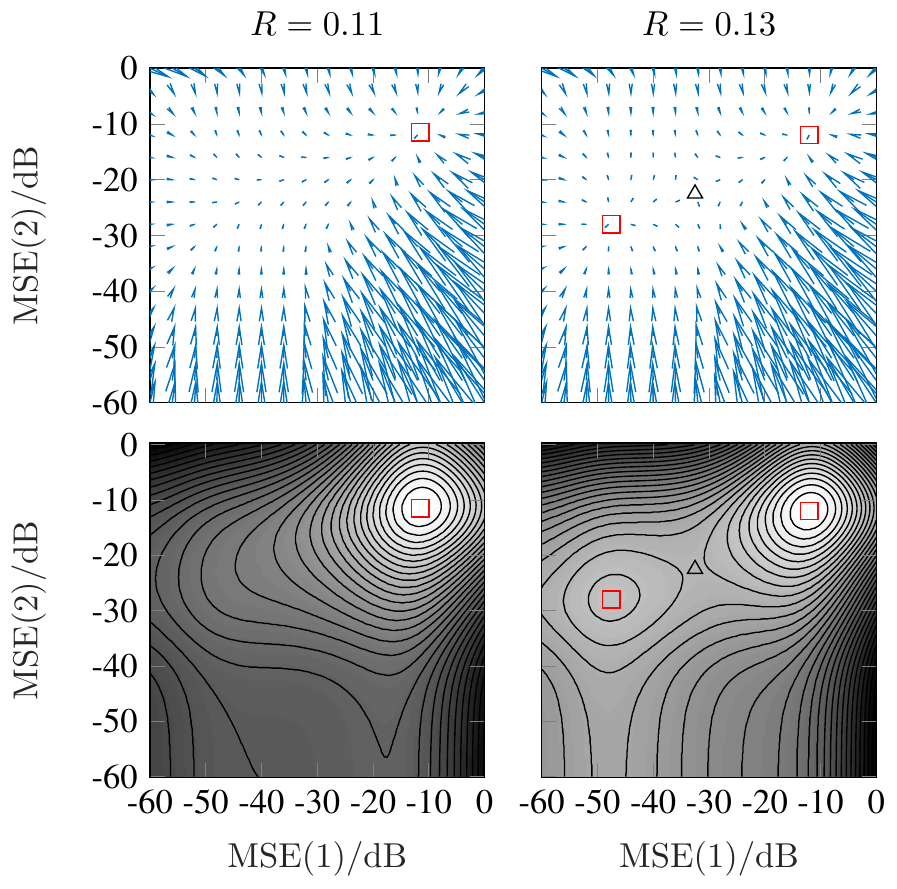}\hspace*{-1mm}
\includegraphics[height=0.48\linewidth]{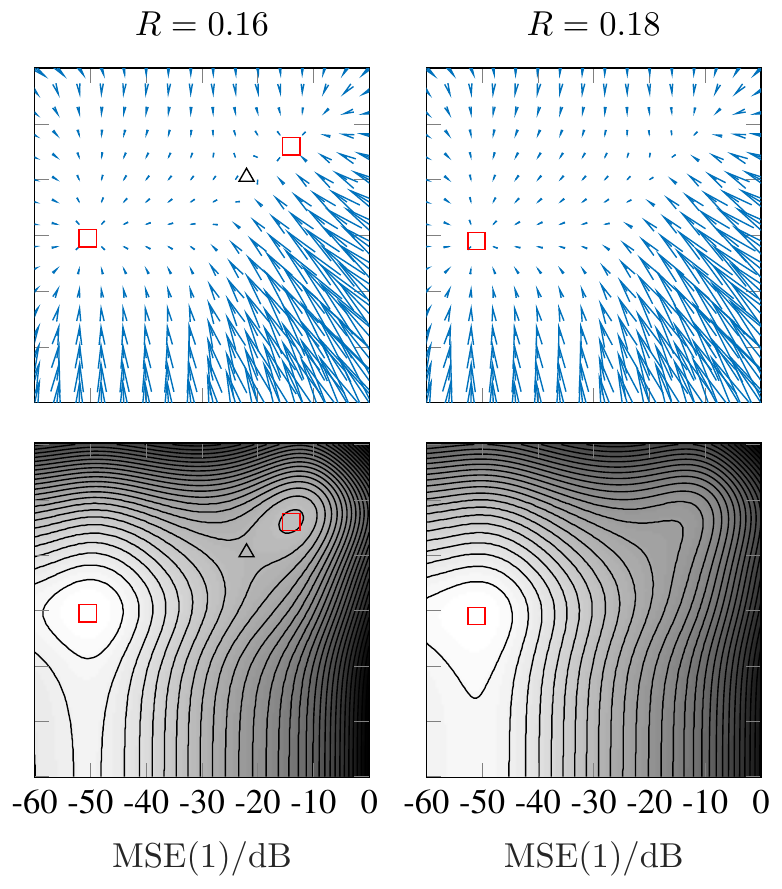}
	\caption{\ac{se} (top) and free energy (bottom) for $B=2$, $\eps=0.1$, $\Sx=\Iv_2$, $\sw^2(1)=-45\dB$, and $\sw^2(2)=-25\dB$. Red squares indicate stable fixed points and local maxima whereas black triangles indicate unstable fixed points and saddle points.} \label{fig:gradientfield_nonuniform}
\end{figure*}

Figure~\ref{fig:gradientfield_nonuniform} illustates an anisotropic scenario with $B=2$ and the channel noise independent but with different variances: $\sw^2(1)=-45\dB$, and $\sw^2(2)=-25\dB$. In the top row the arrows in the MSE plane depict the 
\ac{se} prediction
\begin{equation*}
(\mse^t(1),\mse^t(2)) \rightarrow (\mse^{t+1}(1),\mse^{t+1}(2)).
\end{equation*}
% (the arrows are scaled down for a clearer picture).

The bottom row shows the free energy function (via gray shading and contour lines). Note that the free energy function is no longer symmetric between channels and both the free energy function and the \ac{se} dynamics are nontrivially 2-D. However, it is interesting to note that the stationary points still appear to lie on a globally attracting 1-D submanifold. This raises the question of whether the $B$-dimensional \ac{se} dynamics can be compressed back into a one-dimensional evolution in some way. 

There is also a close match between the fixed points of the \ac{se} and the stationary points of the free energy function, as well as between the \ac{se} arrows and the gradient of the free energy. This match was confirmed in several other numerical experiments. This opens up the possibility for more detailed investigations with different sets of parameters $\sw^2(1),\ldots,\sw^2(B)$ to shed light on the performance regions and dynamics of \ac{Vbamp}. The question arises whether for a given sparsity $\eps$ and measurement rate $R$ there is a diversity function $\theta_{\eps,R}(\sw^2(1),\ldots,\sw^2(B))$ that describes the effective number of jointly sparse measurements based on the individual SNR. More specifically, we expect such a diversity function to combine the \acp{snr} such that, for a certain threshold $B_0$, the global maximum of the free energy equals the \ac{Vbamp} fixed point for $\theta_{\eps,R} \geq B_{0}$ while for $\theta_{\eps,R} < B_{0}$ the free energy has local maxima to the right of the global maximum, which then is no longer the \ac{Vbamp} fixed point. 

% ##################################################################################

\section{Single-Pixel Color Imaging}\label{sec:sin_pix}

We applied MMV-BAMP (cf.~Algorithm \ref{alg:vbamp}) to color imaging using the single-pixel approach from \cite{duarte2008sp}. 
Here, white light illuminates an object and $M$ random 
0/1-masks of dimension $\sqrt{N} \times \sqrt{N}$ with exactly $N/2$ ones are applied before the intensities of the red ($b=1$), green ($b=2$), and blue ($b=3$) components are measured by noisy single-pixel sensors (hence, $B=3$). 
The $B=3$ discrete cosine transform (DCT) coefficient vectors of the acquired image are assumed to be jointly sparse and drawn from a multivariate \ac{bg} pdf (with the exception of the DC term as explained below). The measurement matrix is given by $\mathbf{A} = \mathbf{\Phi}\mathbf{D}^T$,
where the $M \times N$ matrix $\mathbf{\Phi}$ contains the $M$ vectorized binary masks 
and $\mathbf{D}$ is the DCT matrix.
Since $\mathbf{A}$ is the same for all $B=3$ color channels we have an MMV problem.
The measurement matrix $\mathbf{A} $ does not satisfy the conditions 
(zero mean and normalized columns)  required for BAMP. Appendix \ref{app:SinglePixelMeasurementMatrix}
explains how to convert this problem into an equivalent form that meets the BAMP requirements.

\begin{figure}[t]
\centering  
\includegraphics[width=0.95\linewidth]{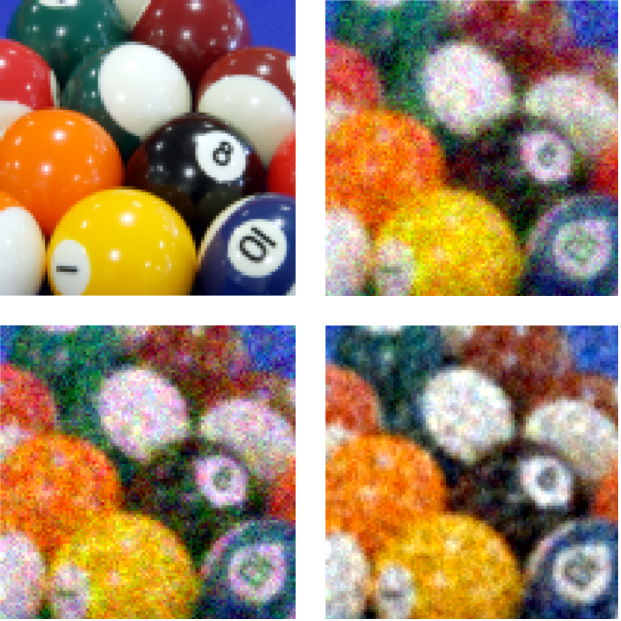}
\vspace*{2mm}
  \caption{Performance comparison for single-pixel color imaging at $R=0.333$:
  original image (top left),
  per-channel AMP with soft-thresholding (top right),
  per-channe BAMP (bottom left),
  and MMV-BAMP (bottom right).
  }
    \label{fig:NumRes1}
\end{figure}

{\subsection{Real-world Data} \label{realworld}}
In order to benchmark the recovery algorithms in a real-world setting, 
we randomly selected a training set of $40$ natural images (see \cite{AsGi:2014,AsGi:2015}) and a distinct 
test image (shown in Figure~\ref{fig:NumRes1}). All images had a resolution of $100 \times 100$ pixels ($N=10\,000$).
The parameters of the BG prior (sparsity $\epsilon$ and covariance matrix $\Sx$) 
and the parameters of the three scalar \ac{bg} priors (one for each color channel) 
were estimated from the training set using the \ac{em} algorithm \cite{bishop2006pattern}. 
% These were fed the scalar (B)AMP algorithm instances that ran independently in each color channel.
The measurement noise was i.i.d.\ zero-mean Gaussian with a standard deviation of $\sw(1) = \sw(3) = 1.5$ for the red and the blue channels
and $\sw(2) = 6$ for the green channel.
The number of measurements was $M=3330$ ($R=0.333$).

Figure~\ref{fig:NumRes1} shows the recovery results for 
(i) AMP with soft thresholding \cite{donoho2009message}, applied independently in each color channel (using the optimal threshold parameter), 
(ii) scalar \ac{bamp}, independently applied in each color channel, and 
(iii) MMV-BAMP (using the estimated \ac{bg} prior). 
Figure \ref{fig:NumRes1} shows that MMV-BAMP indeed outperforms the scalar schemes. 
Since the color channels are affected by different noise variance, 
per-channel AMP and BAMP suffer from a color mismatch.
In contrast, MMV-BAMP does not suffer from this problem
and yields less blurry edges and clearer image details.  

\begin{table}[t]
  \begin{center}
%\normalsize
\vspace*{2mm}
    \caption{Mean recovery NMSE for AMP, BAMP, MMV-BAMP, and group lasso (the 95\%-confidence levels are approximately $\pm1 \dB$). 
    }
            \label{tab:numres1}
\vspace*{2mm}
    \begin{tabular}{lccc}
\toprule
 & \multicolumn{3}{c}{NMSE [dB]} \\ 
 \cmidrule{2-4} 
   & red & green & blue \\
%\midrule
%SNR/dB      &  $64.8$ & $51.9$ & $62.6$\\
\midrule
    AMP      & $-16.5$  & $-12.7$  & $-14.5$  \\
    BAMP     & $-16.6$  & $-12.2$  & $-14.7$ \\
    MMV-BAMP & $-16.8$  & $-14.1$  & $-14.9$ \\
    group lasso & $-16.4$ & $-13.3$ & $-14.5$ \\
       \bottomrule
    \end{tabular}
 \end{center}
\end{table}

Table \ref{tab:numres1} shows the normalized mean square recovery error (NMSE) 
achieved by the various methods on the three color channels (the NMSE was estimated
by averaging over 40 test images). The table also shows the results obtained with
the group lasso \cite{BoydADMM} based on ADMM \cite{BoydADMM,boyd2011distributed3,boyd2011distributed4}
with hand-optimized regularization parameter.

MMV-BAMP is seen to outperform all competing schemes.
Its performance advantage is most pronounced for the green channel, which has the poorest SNR
of $51.9$\,dB.
For the red and blue channels (SNR $64.8$ and $62.6$\,dB, respectively), the performance differences tend to be smaller.
We emphasize that MMV-BAMP achieves these performance gains in spite of a mismatched prior, i.e.,
the distribution of the (jointly sparse) DCT coefficients of natural images is not actually \ac{bg}.

\subsection{Synthetic Data}
To eliminate effects resulting from mismatched priors, 
we next consider artificial images whose red, green, and blue channel
DCT coefficients are jointly sparse and have \ac{bg} distribution.
More specifically, we created images having a resolution of $100 \times 100$ pixels 
($N=10\,000$) by randomly drawing $20\times 20=400$ low-frequency DCT coefficients (on each color channel) from a Gaussian
distribution with covariance matrix $(\Sx)_{ij}=4-|i\!-\!j|$, $i,j\in\{1,2,3\}$ (except the DC coefficients that had a fixed value of $20$). The remaining $9\,600$ high-frequency DCT coefficients per channel were set to zero. The resulting sparsity equals $\epsilon = 400/10\,000 = 4\%$. 

\begin{figure}[t]
\centering
  \includegraphics[width=0.95\linewidth]{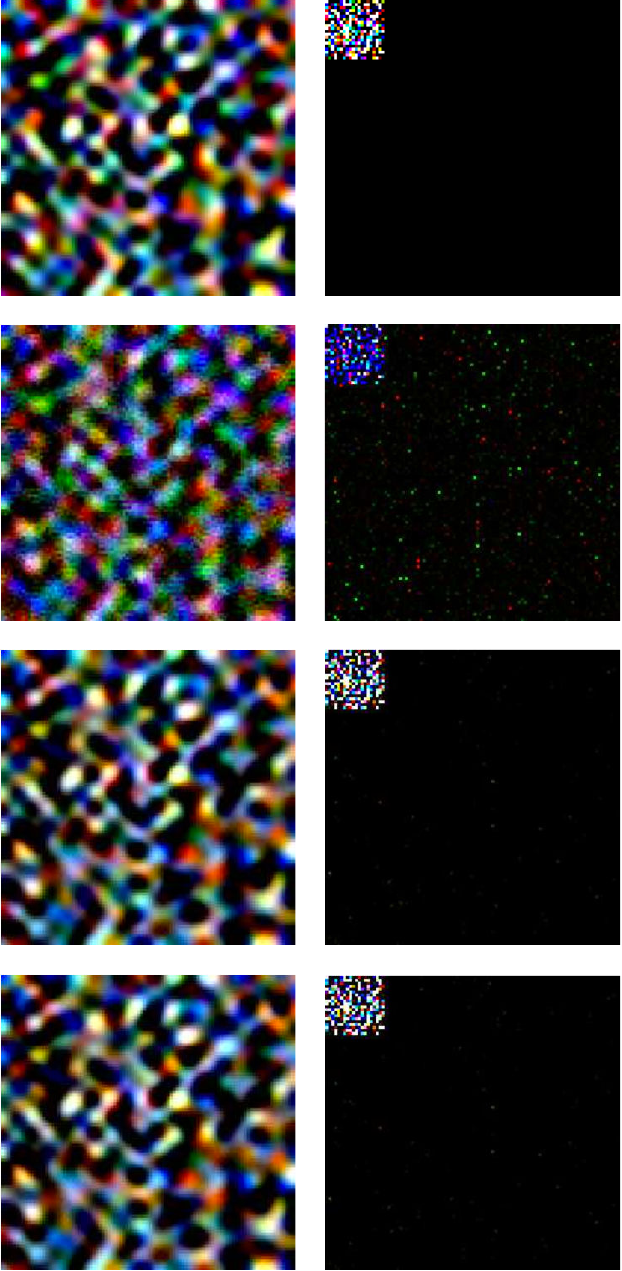}
  \vspace*{2mm}  
  \caption{
  Single-pixel recovery of an artificial image (left column) at $R=0.333$ and corresponding DCT coefficients (right column): original image (top row), BAMP (second row), MMV-BAMP (third row),
  and MMV-BAMP-EM (bottom row).
  \label{fig:NumRes3}}
\end{figure}

We then applied compressive single-pixel imaging as described above 
to these artificial images. The sampling rate was $R=0.333$ and the
standard deviation of the measurement noise 
in the red and the green channels was eight times larger than that in the blue channel
leading to measurement SNRs of $32.4$\,dB, $32.4$\,dB, and $50.5$\,dB, respectively. 
Recovery was done using BAMP, MMV-BAMP with perfect prior knowledge, and a practical variant labeled 
MMV-BAMP-EM. The latter augments MMV-BAMP with an on-the-fly
(i.e., during the recovery iterations) EM-based estimation of the model 
parameters (sparsity, mean, and covariance in the \ac{bg} prior).
As shown in \cite{SchniterEMAMP}, this is possible whenever the
structure of the prior distribution is known. More specifically,
the EM algorithm is applied in Algorithm \ref{alg:vbamp} after 
line 6 to estimate the parameters of a mixture of two multivariate Gaussians 
from the decoupled measurements $\uv^{t-1}(b)$. 
The covariance of the stronger of the two mixture components 
is discounted for the noise and retained for the non-zero part of the \ac{bg} model.

Figure~\ref{fig:NumRes3} shows the results for an exemplary artificial image and its DCT. 
MMV-BAMP is seen to perform much better than BAMP. Furthermore, MMV-BAMP-EM yields 
recovery results virtually identical to MMV-BAMP. Thus, estimating the prior parameters 
during recovery induces a negligible performance loss (indeed, we verified that the EM estimates of
sparsity and covariance were close to the true values even though based on only $3 \times 400$ nonzero DCT 
coefficients). 
The DCT domain results also 
show that the majority of errors occurs in the red and the green channels that suffer
from poor SNR.

\begin{table}[t]
  \begin{center}
    \caption{Mean recovery NMSE for AMP, BAMP, MMV-BAMP, MMV-BAMP EM, and group lasso (the 95\%-confidence levels are less than $\pm 0.1\dB$). 
    }\label{tab:numres2}
\vspace*{2mm}
    \begin{tabular}{lccc}
\toprule
 & \multicolumn{3}{c}{NMSE [dB]} \\ 
 \cmidrule{2-4} 
   & red & green & blue \\
%\midrule
%     measurement SNR/dB       & $32.4$  & $32.4$   & $50.5$\\
\midrule
    AMP           &  $-2.76$ & $-2.81$   & $-17.53$  \\
    BAMP          & $-3.92$  & $-3.96$   & $-23.32$\\
    MMV-BAMP      & $-8.36$  & $-9.52$   & $-23.91$\\
    MMV-BAMP-EM   & $-8.34$  & $-9.49$   & $-23.90$ \\
    group lasso (small $\lambda$)  & $-0.21$ & $-0.30$ & $-11.10$ \\
    group lasso (moderate $\lambda$)   & $-5.00$ &  $-5.30$ & $\;\,-6.80$ \\
       \bottomrule
    \end{tabular}
 \end{center}
\end{table}

A systematic performance comparison in terms of NMSE (obtained by averaging over 100 artificial images)
is provided in Table \ref{tab:numres2}, which also shows the results achieved by the
group lasso. 
It is seen that MMV-BAMP and MMV-BAMP-EM achieve almost identical NMSE and 
outperform (B)AMP 
by exploiting the correlation between the color channels.
The performance gain is specifically noticeable in the low-SNR red and green channels,
with the gain in the green channel being slightly larger since its correlation with
the high-SNR blue channel is stronger ($(\Sx)_{23}=3$)
than that of the red channel ($(\Sx)_{13}=2$). 

The group lasso is seen to perform much worse than MMV-BAMP(-EM) since it 
is unaware of the different measurement SNRs on the three channels.
With weak regularization (small $\lambda$), the group lasso relies more
on the measurements and hence yields reasonable performance only for the high-SNR blue channel. 
With stronger regularization (moderate $\lambda$), the group lasso enforces
stronger sparsity, which is beneficial for the low-SNR red and green channels 
but leads to increased distortions on the blue channel. 
This shows that MMV-BAMP has strong advantages over group lasso when the quality of the measurements of the correlated components is different and unknown. 

% ##################################################################################

\section{Conclusions}\label{sec:concl}

We reviewed the multivariate \ac{Vbamp} algorithm for \ac{mmv}/\ac{dcs} \ac{cs} recovery and its associated multivariate \ac{se}. We established that for arbitrary \ac{mmv} measurement models there is an equivalent model in which signal and noise are both decorrelated. For the widely employed multivariate \ac{bg} signal prior, we proved that uncorrelatedness is preserved during the \ac{Vbamp} and \ac{se} iterations; thus, the complexity of \ac{Vbamp} for \ac{bg} signals scales only linearly with the number of jointly sparse vectors. The free energy formula for the jointly sparse \ac{bg} \ac{cs} channel with $B$ degrees of freedom has been derived and 
juxtaposed with the multivariate \ac{se}. Our results allowed us to assess the impact of signal correlation and of the number of jointly sparse vectors on the phase transition phenomenon and the optimality rate region of \ac{Vbamp}. 
{Numerical results for single-pixel color imaging demonstrated that MMV-BAMP achieves superior recovery quality by exploiting correlation between the vector components. MMV-BAMP can be augmented
with \ac{em}-based estimation of the parameters of the \ac{bg} prior, leading to a practical and flexible scheme with excellent recovery performance and significantly smaller complexity than competing approaches such as group lasso.}

\appendix

\subsection{Equivariance of MMV VBAMP and its SE} \label{app:vbamp_equivariance}

Consider Algorithm \ref{alg:vbamp} with the transformed variables $\Sxt$, $\Tv\xvechat_n^t$, $\Tv\rvec_m^t$, $\Tv\uvec_n^t$, $\Svvt^t$.
Lines $5$ and $6$ are trivially equivariant.
The equivariance of line 7 follows from the invariance property of \ac{mmse} estimators to affine transformations \cite[Ch.~11.4]{kay1993fundamentals}.
In the residual term (line 8), the equivariance of $\yvec_m - \big(\Av(1) \xvhat(1)^t,\ldots,\Av(B) \xvhat(B)^t\big)_m$ is trivial.
It remains to show that the Onsager term is equivariant.
Thus, we write the transformed Onsager term as
\begin{align*}
	&\frac{1}{M} \sum_{n=1}^{N} F'(\Tv \uvec_n ; \Tv\Svv\Tv^T) \Tv \rvec_m \\
	&\stackrel{\{1\}}{=} \frac{1}{M} \sum_{n=1}^{N} \cov\{ \tilde{\xvec} \mid \Tv \uvec_n ; \Tv\Svv\Tv^T \} \big(\Tv\Svv\Tv^T\big)^{-1} \Tv \rvec_m \\
	&= \frac{1}{M} \sum_{n=1}^{N} \Exp\{ \boldsymbol{\langle} \tilde{\xvec} - \Exp\{\tilde{\xvec}\} \boldsymbol{\rangle} \mid \Tv \uvec_n ; \Tv\Svv\Tv^T \} \Tv^{-T} \Svv^{-1} \rvec_m \\
	&= \frac{1}{M} \sum_{n=1}^{N} \Tv \Exp\{ \boldsymbol{\langle} \xvec - \Exp\{\xvec\}\boldsymbol{\rangle} \mid \uvec_n ; \Svv \} \Tv^T \Tv^{-T} \Svv^{-1} \rvec_m \\
	& \stackrel{\{2\}}{=} \Tv \frac{1}{M} \sum_{n=1}^{N}  \cov\{ \xvec \mid \uvec_n ; \Svv \} \Svv^{-1} \rvec_m \\
	& = \Tv \frac{1}{M} \sum_{n=1}^{N} F'(\uvec_n ; \Svv) \rvec_m\,,
\end{align*}
where $\{1\}$ and $\{2\}$ follow from Lemma \ref{lemma1} in Appendix D.
The equivariance of \ac{se} follows by similar arguments using 
elementary probability theory and the invariance property of \ac{mmse} estimators to affine transformations \cite[Ch.~11.4]{kay1993fundamentals}. %, 2) can be proven by simple calculation.

\subsection{Diagonality of SE with BG Prior} \label{app:bg_diagonality}

We show that \ac{mmv} \ac{se} \equref{eq:stateevolution} preserves diagonality for the \ac{bg} prior.
In particular, we prove that if $\Svv^{t}$, $\Sw$ and $\Sx$ are diagonal, then 
\begin{equation*}
\Svv^{t+1} = \Sw + \frac{1}{R} \underbrace{\Exp_{\xvecs,\vvecs} \lef\{ \boldsymbol{\langle} F(\xvecs+\vvecs^t;{\Svv}^{t}) - \xvecs\boldsymbol{\rangle}  \rig\} }_{\Cv}
\end{equation*}
is also diagonal. It suffices to establish that $\Cv$ is diagonal.
% Since the factor $\frac{1}{R}$ and the term $\Sw$ do not influence the diagonality, we examine the expectation $\Cv$.
Inserting the \ac{bg} prior \equref{eq:signal_prior} and its estimator \equref{eq:2_Fvec_bernoulli_gaussian} and writing out the integrals for $(\Cv)_{i,j}$ ($i,j=1,\ldots,B$), it is seen that for $i\neq j$ the integrands have odd symmetry 
% around $0$) 
w.r.t.\ a separable set of their arguments and thus integrate to $0$. 
It follows that $(\Cv)_{i,j}=0$ for $i\neq j$ and that $\Svv^{t+1}$ is diagonal.
%{Or should we insert a short appendix for this proof?}

\subsection{Replica Analysis} \label{app:replica}

Following the analysis in \cite{zhu2016performance}, we derive an analytical performance prediction for the \ac{Vbamp} algorithm for \ac{mmv} and \ac{dcs} problems.
We consider the measurement model \eqref{eq:matrixmeasurement}
and the signal prior \eqref{eq:signal_prior} with $\Sx=\Iv$
and 
$\wvecs_m \sim \mcal{N}(\zerov,\Sw)$, where $\Sw= \diag(\sw^2(1),\ldots\sw^2(B))$ is a diagonal matrix with the noise variances $\sw^2(b)$.
The special case $\Sw = \sw^2 \Iv$ was analyzed in \cite{zhu2016performance}.
We follow \cite{zhu2016performance} by assuming the rows of $\Avs(b)$
% , i.e., $(\Avs(b))_{m,n}$ 
to have variance $\frac{1}{N}$.
The straightforward rescaling to normalized columns is discussed at the end.
For the sake of notational simplicity, the following derivation applies to the \ac{mmv} scenario, i.e., $\Avs(1)=\ldots=\Avs(B)=\Avs$.
The generalization to \ac{dcs} is straightforward (cf.\ \cite{zhu2016performance}).
The posterior \ac{pdf} of the estimate $\Xvhat = (\xvhat(1),\ldots,\xvhat(B)) = (\xvechat_1,\ldots,\xvechat_N)^T$ reads
\begin{equation*}
f_{\Xvhat \mid \Yvs}(\Xvhat \mid \Yv) = \frac{1}{Z} \prod_{n=1}^{N} f_{\hat{\xvecs}}(\xvechat_n) \prod_{m = 1}^{M} \mcal{N}\big((\Yv\!-\!\Av\Xvhat)_m; \zerov, \Sw \big)
\end{equation*}
with $\Yv = (\yv(1),\ldots,\yv(B)) = (\yvec_1,\ldots,\yvec_M)^T$. 
Furthermore, $Z$ is the partition function
\begin{equation*}
Z =\! \int\displaylimits_{\mbb{R}^{N B}} 
\! \prod_{m=1}^{M}\! \mcal{N}\big((\Yv\! - \!\Av\Xvhat)_m; \zerov, \Sw \big) 
\!\prod_{n=1}^{N} \!f_{\hat{\xvecs}}(\xvechat_n) \,d\xvechat_n. 
% \prod_{n=1}^{N} d\xvechat_n \,. 
\end{equation*}
Following the argumentation in \cite{zhu2016performance} and the assumptions in \cite{tanaka2002statistical,guo2005randomly,krzakala2012probabilistic,krzakala2012statistical,mezard2009information,barbier2015approximate}, we determine the stationary points
%fixed points (local maxima)
of the free energy function, which provide the \acp{mse} in the fixed points of \ac{Vbamp} along with the \ac{mmse} for the measurement model \equref{eq:matrixmeasurement}.
The free energy is defined as 
\be\label{freeenergy_replica}
\mcal{F} = \lim_{N\rightarrow\infty} \frac{1}{N} \Exp_{\Avs,\Xvs,\Wvs} \lef\{ \log(Z) \rig\} ,
\ee
but in general is difficult to evaluate.
The replica method \cite{tanaka2002statistical,guo2005randomly,krzakala2012probabilistic,krzakala2012statistical,mezard2009information,barbier2015approximate} introduces $k$ replicas $\Xvhat^1,\ldots,\Xvhat^k$ of the estimate $\Xvhat$ and 
approximates the free energy \equref{freeenergy_replica} as
\begin{equation}
\mcal{F} = \lim_{N\rightarrow\infty} \lim_{k\rightarrow 0} \frac{ \Exp_{\Avs,\xvs,\wvs} \lef\{ Z^k \rig\} -1 }{Nk} \,. \label{replicatrick}
\end{equation}
The self-averaging property that leads to \equref{freeenergy_replica} and the replica trick \equref{replicatrick} as well as the replica symmetry assumptions are assumed to be valid,
even though their theoretical justification is still an open problem % in mathematical physics 
\cite{tanaka2002statistical,guo2005randomly,krzakala2012probabilistic,krzakala2012statistical,mezard2009information,barbier2015approximate}.
In order to evaluate \equref{freeenergy_replica}, we write
\begin{align}\label{bigexpectation_replica}
&\Exp_{\Avs,\xvs,\wvs} \!\lef\{ Z^k \rig\} \!=\! 
\lvert 2\pi \Sw \rvert^{-\frac{k}{2}} 
% \nonumber \\ &\phantom{=} 
\Exp_{\Xvs} \!\bigg\{ \!\int %\displaylimits_{\mbb{R}^{NB}} 
	\!\prod_{m=1}^{M} \!\mbb{X}_m 
	\!\prod_{n=1}^{N}\! \prod_{a=1}^{k} \!f_{\xvecs}(\xvec_n^a) 
% \prod_{n=1}^{N} \prod_{a=1}^{k} 
\,d\xvec_n^a  \!\bigg\} ,
\end{align}
where
\begin{equation}\label{vmexp_replica}
% \mbb{X}_m = \Exp_{\Avs,\Wvs} \!\lef\{ \exp\!\lef( \!-\frac12 \sum_{a=1}^{k} \sum_{b=1}^{B} \frac{1}{\sw^2(b)}(v_{m,b}^a)^2 \!\rig) \!\rig\} ,
\mbb{X}_m = \Exp_{\Avs,\Wvs} \!\Big\{\! \exp\!\Big( \!-\frac12  \| \bar{\vvec}_m \|^2 \!\Big) \!\Big\} .
\end{equation}
Here, we used the vector 
$\bar{\vvec}_m = \bar{\boldsymbol{\Sigma}}_{\wvecs}^{-\frac12} \vvec_m $
defined in terms of 
$\vvec_m = (v^1_{m,1},\ldots,v^k_{m,1},v^1_{m,2},\ldots,\ldots,v^k_{m,B})^T$,
and
$\bar{\boldsymbol{\Sigma}}_{\wvecs} = \Sw \otimes \Iv_{k\times k}$, 
where the elements of $\vvec_m$ are in terms of
\begin{equation*}
\vvec_{m}^a = (v_{m,1}^a,\ldots,v_{m,B}^a) =   \big(\Av (\Xv \!-\! \Xvhat^a) + \Wv\big)_m .
\end{equation*}

Using a Gaussian approximation for the \ac{pdf} of $\bar{\vvec}_m$, 
\be\label{vmdistribution_replica}
f_{\bar{\vvecs}_m}\lef(\bar{\vvec}_m\rig) = \mcal{N}\lef(\bar{\vvec}_m;\zerov,\Gv_m\rig) ,
\ee
\equref{vmexp_replica} can be evaluated as % and \equref{vmdistribution_replica} one obtains
\begin{align}\label{wq:Xmdet}
 \mbb{X}_m =
 % & = \Exp_{\bar{\vvec}_m} \left\{\exp\lef( -\frac12 \bar{\vvec}_m  \bar{\vvec}_m^T \rig)\right\}\\ &= 
 \lvert \Iv + \Gv_m \rvert^{-\frac12} .
\end{align}
Here, we used the covariance matrix
$\Gv_m = \cov\{ \bar{\vvec}_m \} = \bar{\boldsymbol{\Sigma}}_{\wvecs}^{-\frac12} \bar\Gv_m \bar{\boldsymbol{\Sigma}}_{\wvecs}^{-\frac{T}{2}}$
with $\Gv_m = \cov\{ {\vvec}_m \} $.
The matrix $\bar\Gv_m$ is composed of $B\times B$ blocks of size $k\times k$ as follows:
\begin{enumerate}
	\item The main diagonal of $\bar\Gv_m$ consists of entries $g_{1}(b) = \Exp_{\Avs,\wvs}\{(v_{m,b}^a)^2\}$, which is different in each of the $B$ blocks but identical within a block.
	\item The remaining entries in the blocks of the main diagonal are $g_{2}(b)=\Exp_{\Avs,\wvs}\{ v_{m,b}^a v_{m,b}^{a'}\}$, which are different in each block but identical within a block.
	\item The diagonal entries of the off-diagonal blocks are $g_3(b,b') = \Exp_{\Avs,\wvs}\{v_{m,b}^a v_{m,b'}^{a}\}$.
	\item The off-diagonal entries of the off-diagonal blocks are $g_4(b,b') = \Exp_{\Avs,\wvs}\{v_{m,b}^a v_{m,b'}^{a'}\}$.
\end{enumerate}
Using the normalization of the measurement matrix $\Av$, the fact that $\xvec_n^a$ follows the same distribution as $\xvec_n$, 
and the replica symmetry \cite{krzakala2012probabilistic,krzakala2012statistical},
these values turn out to be
\begin{align*}
	g_{1}(b) &= \frac{1}{N}\sum_{n=1}^{N} (x_n(b) - \xhat_n^a(b))^2 + 1, \\
	g_{2}(b) &= \frac{1}{N}\sum_{n=1}^{N} (x_n(b) - \xhat_n^a(b))  (x_n(b) - \xhat_n^{a'}(b)) + 1, \\
	g_3(b,b') &= \frac{1}{N}\sum_{n=1}^{N} (x_n(b) - \xhat_n^a(b))  (x_n(b') - \xhat_n^{a'}(b)) , \\
	g_4(b,b') &= \frac{1}{N}\sum_{n=1}^{N} (x_n(b) - \xhat_n^a(b'))  (x_n(b) - \xhat_n^{a'}(b')) .
\end{align*}
By introducing the auxiliary quantities
\begin{align*}
	m_a(b,b') &= \frac{1}{N} \sum_{n=1}^{N} \xhat_n^a(b) x_n(b'), \\ 
	Q_a(b,b') &= \frac{1}{N} \sum_{n=1}^{N} \xhat_n^a(b) \xhat^a_n(b'), \\
	q_{aa'}(b,b') &= \frac{1}{N} \sum_{n=1}^{N} \xhat_n^a(b) \xhat^{a'}_n(b'), \\ 
	q_0(b,b') & = \frac{1}{N} \sum_{n=1}^{N} x_n(b) x_n(b') ,
\end{align*}
the covariance values can be written as
\begin{align*}
	g_{1}(b) &= \eps - 2m_a(b,b) + Q_a(b,b) + 1, \\
	g_{2}(b) &= \eps - m_a(b,b) - m_{a'}(b,b) + q_{aa'}(b,b) + 1,\\
    g_3(b,b') &= {	q_0(b,b') - m_a(b',b) - m_{a'}(b,b) + q_{aa'}(b,b),}\\
    g_4(b,b') &= {	q_0(b,b) - m_a(b',b) - m_{a'}(b',b) + q_{aa'}(b',b') }.
\end{align*}

In the Bayesian setting the distribution of $\xvec_n$ matches the distribution of $\xvechat_n$ and that of the replicas $\xvechat^a_n$, thus $g_3(b,b') = g_4(b,b') = 0$.
Furthermore, due to the replica symmetry \cite{krzakala2012probabilistic,krzakala2012statistical} $m_a(b,b)=m_{a'}(b,b)=m(b)$, $Q_a(b,b) = Q(b)$, and $q_{aa'}(b,b)=q(b)$.
It follows that the $\Gv_m$ is a structured matrix that, due to its block structure, can be expressed in terms of all-ones matrices, identity matrices, and Kronecker products. Its $kB$ eigenvalues can straightforwardly be determined as
\begin{align*}
	\alpha^b_1 = g_{1}(b) + (k\!-\!1)g_{2}(b), %  &\times& 1,\quad &b=1,\ldots,B\,,\\
	\qquad
	\alpha^b_2 = g_{1}(b) - g_{2}(b), 
	% &\times& (k-1),\quad &b=1,\ldots,B \,.
\end{align*}
where the $\alpha^b_1$ have multiplicity 1 and the $\alpha^b_2$ have multiplicity $k\!-\!1$.
We can thus express \eqref{wq:Xmdet} as
\begin{align*}
	\lvert \Iv + \Gv_m \rvert^{-\frac12} &= \Bigg[ \prod_{b=1}^{B} \Big( 1+k\frac{\eps - 2m(b) + q(b) + \sw^2(b)}{\sw^2(b)+Q(b)-q(b)} \Big) \\ 
	& \phantom{=} \prod_{b=1}^{B}\Big( 1+ \frac{1}{\sw^2(b)}(Q(b)-q(b)) \Big)^{k-1} \Bigg]^{-\frac12} \,.
\end{align*}
Using the Taylor series approximation
\begin{equation}\nonumber
% \exp(x) \approx 1+x \Rightarrow (1+x)^{-\frac12} \approx \exp\lef(-\frac{x}{2}\rig)
\exp\lef(-\frac{x}{2}\rig) \approx (1+x)^{-\frac12} ,
\end{equation}
we obtain
\begin{align*}
\lim_{k\rightarrow 0} \mbb{X}_m &= \exp \Big( -\frac{k}{2}\sum_{b=1}^{B} \frac{\eps-2m(b)+q(b)+\sw^2(b)}{\sw^2(b)+Q(b)-q(b)} \\
& \phantom{=} - \log(Q(b)-q(b)+\sw^2(b)) - \log(\sw^2(b)) \Big)\,.
\end{align*}
Following the derivation in \cite[App.]{zhu2016performance}, \equref{bigexpectation_replica} can be written as
\scriptsize
\begin{equation*}
\Exp_{\Avs,\Xvs,\Wvs} \lef\{ Z^k \rig\} = \int \exp \lef( kN\Phi(m_0,\hat{m}_0,q,\hat{q},Q,\hat{Q}) \rig) dm_0\,d\hat{m}_0\,dq\,d\hat{q}\,dQ\,d\hat{Q}\,.
\end{equation*}
\normalsize
Remember that we are only interested in the stationary points of the free energy expression \equref{bigexpectation_replica}.
Thus, we set
\scriptsize
	\begin{align}
	&\mcal{F} = \Phi(\{m(b)^*,\hat{m}(b)^*,q(b)^*,\hat{q}(b)^*,Q(b)^*,\hat{Q}(b)^*\}_{b=1,\ldots,B})\nonumber \\
	&  = \frac{1}{2}\sum_{b=1}^{B}\lef(Q(b)\hat{Q}(b) - 2m(b)\hat{m}(b) + q(b)\hat{q}(b)\rig) -\frac{R}{2}\log\big(\lvert 2\pi\Sw \rvert\big)\nonumber\\
	& - \frac{R}{2}\sum_{b=1}^{B} \Bigg( \frac{\eps-2m(b)+q(b)+\sw^2(b)}{Q(b)-q(b)+\sw^2(b)} \nonumber\\
	&\hspace{15mm}+ \log\big(Q(b)-q(b)+\sw^2(b)\big) - \log\big(\sw^2(b)\big) \Bigg)\nonumber\\
	& +\int_{\mbb{R}^B} f_{\xvec}(\xvec) \int_{\mbb{R}^B} \log \int_{\mbb{R}^B} f_{\xvechat}(\xvechat) \nonumber\\
	& \prod_{b=1}^{B}\exp\Big(-\frac12\hat{q}(b)\,\xhat(b)^2+\hat{m}(b)\,\xhat(b)x(b)+\sqrt{\hat{m}(b)}\,\xhat(b)h(b)\Big) d\xvechat\, \mcal{D}\hvec\, d\xvec , \label{freeenergy36}
	\end{align}
\normalsize
where the superscript $\cdot^*$ denotes stationary points.
The stationary points are obtained by differentiation as
\small
\begin{align*}
	\frac{d\Phi}{dm(b)}&=0 \, \Rightarrow \, \hat{m}(b)^* = \frac{R}{E(b)+\sw^2(b)} = \gamma(b), \\
	\frac{d\Phi}{dq(b)}&=0 \, \Rightarrow \, \hat{q}(b)^* = \frac{R}{E(b)+\sw^2(b)} = \gamma(b), \\
	\frac{d\Phi}{dQ(b)}&=0 \, \Rightarrow \, \hat{Q}(b)^* = 0.
\end{align*}
\normalsize
Here, we used the substitution $E(b)=Q(b)-q(b)$, and the fact that in the Bayesian setting $q(b)^*=m(b)^*$, and $Q(b)^*=\eps$.
Substituting back into \equref{freeenergy36} and using $\Evec=(E(1),\ldots,E(B))^T$, we obtain
\scriptsize
	\begin{align*}
	&\mcal{F}(\Evec,\Sw)  =\\ &-\frac{R}{2}\sum_{b=1}^{B}\lef(\log\big(2\pi(\sw^2(b)+E(b))\big)+\frac{\eps+\sw^2(b)}{E(b)+\sw^2(b)} \rig) \\
	&\phantom{=} + \int_{\mbb{R}^B}   f_{\xvec}(\xvec) \int_{\mbb{R}^B} \log \Bigg( \int_{\mbb{R}^B} f_{\xvechat}(\xvechat)\\
	&\phantom{=} \prod_{b=1}^{B}\exp\Big(-\frac12\gamma(b)\,\xhat(b)^2+\gamma(b) \xhat(b)x(b)+\sqrt{\gamma(b)} \xhat(b)h(b)\Big) d\xvechat \Bigg)\, \mcal{D}\hvec\, d\xvec,
	\end{align*}
\normalsize
where the second integration is over a standard Gaussian measure, i.e., $\mcal{D}\hv = \prod_{b=1}^{B} \mcal{N}(h_b;0,1)dh_b=\mcal{N}(\hv;\zerov,\Iv)\prod_{b=1}^{B}dh_b$.
Inserting the signal prior \equref{eq:signal_prior} results in
\scriptsize
\begin{align*}
&\mcal{F}(\Evec,\Sw) =-\frac{R}{2}\sum_{b=1}^{B}\lef(\log\lef(2\pi(\sw^2(b)+E(b))\rig)+\frac{\eps+\sw^2(b)}{E(b)+\sw^2(b)}\rig) \\
&  +	(1-\eps)\int \log\Bigg( (1-\eps) + \\
& \eps\int\exp\big(-\frac12\gamma(b) \xhat^2+\sqrt{\gamma(b)} \xhat(b)h(b)\big) \mcal{D}\xv \Bigg)\mcal{D}\hvec \\
&  + \eps \int \int \log\Bigg( (1-\eps) + \\
&  \eps\int \exp\big(-\frac12\gamma(b)\xhat(b)^2 + \gamma(b)\xhat(b) x(b) + \sqrt{\gamma(b)} \xhat(b) h(b)\big)  \mcal{D}\xvechat \Bigg) \mcal{D}\hvec\,\mcal{D}\xvec \,,
\end{align*}
\normalsize
with the measures $\mcal{D}\xvec$ and $\mcal{D}\xvechat$ analogously as above.
Further simplification leads to
\scriptsize
\begin{align*}
&\mcal{F}(\Evec,\Sw) \nonumber\\
&= -\frac{R}{2} \sum_{b=1}^{B}\lef(\log\lef(2\pi(\sw^2(b)+E(b))\rig)+\frac{\eps+\sw^2(b)}{E(b)+\sw^2(b)} - \frac{\gamma(b)(1+\eps\gamma(b))}{R(1+\gamma(b))}\rig) \nonumber\\
&+ (1-\eps)\!\int\! \log \left( \eps \prod_{b=1}^{B}(1+\gamma(b))^{-\frac12} \!+\! (1-\eps) \exp\big(\!-\!\frac{1}{2} \sum_{b=1}^{B} \gamma(b)h^2(b)\big)\! \right) \!\mcal{D}\hvec \nonumber\\
&+ \eps\! \int\! \log \left( \eps \prod_{b=1}^{B}(1+\gamma(b))^{-\frac12} \!+\! (1-\eps) \exp\big(\!-\!\frac12 \sum_{b=1}^{B} \frac{\gamma(b)}{1+\gamma(b)}  h^2(b)\big) \!\right)\!\mcal{D}\hvec  .
\end{align*}
\normalsize
In order to arrive at \equref{eq:freeenergy} that is valid for measurement matrices with normalized columns we use the equivalence between the measurement models with normalized rows and normalized columns and replace $\sw^2(b)$ with $R\sw^2(b)$:
\begin{align*}
\yv = \Av \xv + \wv \quad \Longrightarrow \quad 
\bar{\yv} = \frac{1}{\sqrt{R}} \yv &= \bar{\Av} \xv 
%	+ \frac{1}{\sqrt{R}} \wv \\ = \bar{\yv} &= \bar{\Av} \xv 
	+ \bar{\wv},
\end{align*}
where $\bar{\Av}$ has normalized columns and $\bar{\ws}_m \sim \mcal{N}(0,\frac{\sigma_{\ws}^2}{R})$ if $\ws_m \sim \mcal{N}(0,\sigma_{\ws}^2)$. 

\subsection{Estimator Derivative and Conditional Correlation} \label{app:lemma1}
\begin{lemma}\label{lemma1}
	Given a realization $\xv$ of a random vector $\xvs \in \mbb{R}^N$ with pdf $f_{\xvs}(\xv)$ and its noisy observation
	\begin{equation}\nonumber
	\uv = \xv + \wv 
	\end{equation}
	with $\wvs \sim \mcal{N}(0,\Sw)$ being independent additive Gaussian noise, its \ac{mmse} estimator is
	\begin{equation}\nonumber
	\hat{\xv}(\uv) = \Exp \left\{ \xvs \mid \uvs = \uv  \rig\} \,.
	\end{equation}
	Then, the following relation holds:
	\begin{equation*}
	\cov\lef\{ \xvs \mid \uvs = \uv  \rig\} = \frac{d}{d\uv^T}\hat{\xv}(\uv) \Sw \,.
	\end{equation*}
	\begin{proof}
		Given the definition of the conditional mean and covariance,
		{\small
			\begin{align*}
			\Exp\lef\{\xvs \mid \uv,\Sw \rig\} &= \frac{1}{f_{\uvs}(\uv)} \int_{\mbb{R}^N} \xv f_{\uvs\mid\xvs}(\uv\mid\xv) f_{\xvs}(\xv) d\xv \\
			\cov\lef\{ \xvs \mid \uv,\Sw \rig\} &= \frac{1}{f_{\uvs}(\uv)} \int_{\mbb{R}^N} \xv \xv^T f_{\uvs\mid\xvs}(\uv\mid\xv) f_{\xvs}(\xv) d\xv \\
			&\phantom{=} - \Exp\lef\{\xvs \mid \uv \rig\} \Exp\lef\{\xvs \mid \uv \rig\}^T \,,
			\end{align*}}
		we have
		{\small
			\begin{align}
			\frac{d}{d\uv} \hat{\xv}(\uv) \Sw  =& \frac{1}{f_{\uvs}(\uv)} \int_{\mbb{R}^N} \xv f_{\xvs}(\xv) \frac{d}{d\uv^T}f_{\uvs\mid\xvs}(\uv\mid\xv)  d\xv\, \Sw \nonumber \\
			&\hspace*{-20mm} - \int_{\mbb{R}^N} \frac{1}{f_{\uvs}(\uv)}\xv f_{\uvs\mid\xvs}(\uv\mid\xv) f_{\xvs}(\xv) d\xv \,\, \frac{1}{f_{\uvs}(\uv)} \frac{d}{d\uv^T}f_{\uvs}(\uv)\, \Sw \label{A1}\,.
			\end{align}}
		Since $f_{\uvs \mid \xvs}(\uv\mid \xv) = \mcal{N}(\zerov,\Sw)$ \cite{petersen2008matrix}, 
			\begin{equation}\label{A2}
			\frac{d}{d\uv^T} f_{\uvs \mid \xvs}(\uv\mid \xv) = f_{\uvs \mid \xvs}(\uv\mid \xv) (\xv - \uv)^T \Sw^{-1}  \,.
			\end{equation} 
		Furthermore, the \ac{mmse} estimator can be written as \cite{raphan2007empirical,raphan2011least}
		{\small
			\begin{equation}\label{A3}
			\xvhat(\uv) = \uv + \Sw \frac{1}{f_{\uvs}(\uv)} \frac{d}{d\uv} f_{\uvs}(\uv) \,.
			\end{equation}}
		Combining \equref{A1}, \equref{A2}, and \equref{A3} we have
		{\small
			\begin{align*}
			&\frac{d}{d\uv} \xvhat(\uv) \Sw = \frac{1}{f_{\uvs}(\uv)} \int_{\mbb{R}^N} \xv f_{\uvs \mid \xvs} (\xv \mid \uv) (\xv - \uv)^T f_{\xvs}(\xv) d\xv \\
			&\phantom{=} - (\xvhat(\uv) - \uv) \frac{1}{f_{\uvs}(\uv)} \int_{\mbb{R}^N} \xv f_{\uvs\mid\xvs}(\uv\mid\xv) f_{\xv}(\xv) d\xv \\
			& = \frac{1}{f_{\uvs}(\uv)} \int_{\mbb{R}^N} \xv\xv^T f_{\uvs\mid\xvs}(\uv\mid\xv) f_{\xvs}(\xv) d\xv - \xvhat(\uv) \xvhat(\uv)^T \\
			& = \cov\lef\{ \xvs \mid \uv \rig\} \,,
			\end{align*}}
		which completes the proof.
	\end{proof}
\end{lemma}

%%%%%%%%%%%%%%%%%%%%%%%%%%%%%%%%%%%%%%%%%%%%%%%%%%%%%%

{\subsection{Measurement Conversion for Single-Pixel Imaging}\label{app:SinglePixelMeasurementMatrix}}
We start from the measurement equation (\ref{eq:matrixmeasurement}), where  $\Av(b) = \Av = \mathbf{\Phi} \mathbf{D}^T$, $b=1,2,3$. Hence, 
\begin{equation*}
 \yv(b) = \mathbf{\Phi} \mathbf{D}^T   \xv(b) + \wv(b) \: ,
\end{equation*}
where $\xv(b)$, $b=1,2,3$, are the vectorized DCT coefficients of the red, green, and blue channels, respectively. The $M\times N$ matrix $\mathbf{\Phi}$ 
consists of the $M$ vectorized $0/1$ masks $\boldsymbol{\phi}_i^T$ of dimension $1\times N$, each having exactly $N/2$ ones. 
Furthermore, $\mathbf{D}$ is an $N\times N$ (combined row-column) DCT matrix. 

Since all rows of $\mathbf{\Phi}$ have exactly $N/2$ ones and all elements of the first column of 
$\mathbf{D}^T$ equal $1/\sqrt{N}$, it follows that all elements of the first column $\mathbf{a}_1$ of $\Av$ 
are equal to ${\sqrt{N}}/{2}$
and hence $\mathbf{a}_1$ has mean $\sqrt{N}/2$ and Euclidean norm $\sqrt{MN}/2$.
Since the remaining columns $\mathbf{a}_2,...,\mathbf{a}_N$ of $\mathbf{A}$ 
equal the sum of randomly sampled cosine sequences, their mean is approximately zero
and their norm approximately equals $\lVert\mathbf{a}_i\rVert_2 \approx {\sqrt{M}}/{2}$.
Since BAMP requires a measurement matrix with zero-mean and unit-norm columns,
we compensate for the first column and renormalize the remaining columns, i.e.,
\begin{equation*}
  \underbrace{\frac{\yv(b) -  \mathbf{a}_1 \sqrt{N} x_1(b)}{\sqrt{M}/2}}_{\displaystyle \mathbf{\tilde y}(b)} = \underbrace{\frac{\Big(\mathbf{a}_2,  \mathbf{a}_3, ...,  \mathbf{a}_N \Big)}{\sqrt{M}/2}}_{
\displaystyle \mathbf{\tilde A}  }
    \begin{pmatrix}
    x_2(b)\\
    x_3(b)\\
    ...\\
    x_N(b)
   \end{pmatrix}  + \frac{\wv(b)}{\sqrt{M}/2}  .
\end{equation*}
The new measurement matrix $\mathbf{\tilde A}$ now satisfies the BAMP requirements. It remains to
find the DC coefficients $x_1(b)$, $b=1,2,3$.
Denoting the color component vectors by $\bar{\xv}(b) = \mathbf{D}^T \xv(b)$, we have
$x_1(b) = \sum_{n=1}^N\bar{x}_n(b)/\sqrt{N} $.
Furthermore, since half of the elements of the masks $\boldsymbol{\phi}_i^T$ equal 1 we have
$\sum_{i=1}^M \boldsymbol{\phi}_{i}^T\approx \mathbf{1}M/2$ and hence
\begin{align*}
  \frac{1}{M}\! \sum_{i=1}^{M} y_i(b) = \frac{1}{M}\!
  \sum_{i=1}^{M} \boldsymbol{\phi}_i^T  \bar{\xv}(b) +\!
  \underbrace{\frac{1}{M}\! \sum_{i=1}^{M} w_i(b)}_{\approx 0} %\\ &
  \approx \frac{1}{2} \! \sum_{n=1}^{N}\bar{x}_n(b), 
\end{align*}
thus finally leading to the estimate
\[
x_1(b) \approx \frac{2}{M\sqrt{N}}\sum_{i=1}^{M} y_i(b).
\]

%%%%%%%%%%%%%%%%%%%%%%%%%%%%%%%%%%%%%%%%%%%%%%%%%%%%%%

%%%%%%%%%%%%%%%%%%%%%%%%%%%%%%%%%%%%%%%%%%%%%%%%%%%%%%

%\bibliographystyle{ieeetr}
\bibliography{work_gabor}
%\newpage
%\input{bio/bio_ghannak.tex}
%\input{bio/bio_aperelli.tex}
%\input{bio/bio_ngoertz.tex}
%\input{bio/bio_gmatz.tex}
%\input{bio/bio_mdavies.tex}

% that's all folks
\end{document}